\documentclass[conference]{IEEEtran}
\IEEEoverridecommandlockouts
\usepackage{float}
\usepackage{cite}
\usepackage{array}
\usepackage{fvextra}
\usepackage{amsmath,amssymb,amsfonts}
\usepackage{algorithm}
\usepackage{graphicx}
\usepackage{textcomp}
\usepackage{xcolor}
\usepackage{soul}
\usepackage{algorithm}
\usepackage[hyphens]{url}
\usepackage{algpseudocode}
\usepackage{booktabs}
\usepackage{multirow}
\usepackage{makecell} 
\usepackage{tabularx}
\usepackage{ifthen}
\usepackage{pifont}
\usepackage{ragged2e}
\usepackage{enumitem}
\usepackage{tabularray}
\usepackage{tablefootnote} 
\usepackage{threeparttable}
\usepackage{placeins}
\usepackage[hidelinks]{hyperref}
\IfFileExists{ulem.sty}{\usepackage[normalem]{ulem}}{\providecommand{\sout}[1]{\st{#1}}}
\definecolor{RevBrown}{RGB}{200,150,100}
\definecolor{lightercyan}{RGB}{224,238,238} 
\definecolor{darkgreen}{RGB}{0,180,0}
\definecolor{RevDarkRed}{RGB}{139,0,0}   
\definecolor{lightblue}{rgb}{0.68, 0.85, 0.9} 
\definecolor{lightgreen}{rgb}{0.56, 0.93, 0.56} 
\definecolor{lightorange}{rgb}{1, 0.8, 0.6} 
\definecolor{lightpurple}{rgb}{0.91, 0.74, 0.91} 
\definecolor{lightbrown}{RGB}{231,196,156}
\definecolor{yellow}{rgb}{0.94, 0.94, 0.6} 
\definecolor{ShepherdBlue}{RGB}{0,92,184}

\newif\ifshowshepherdchanges
\showshepherdchangesfalse
\ifshowshepherdchanges
  \newcommand{\shepdel}[1]{\sout{#1}}
  \newcommand{\shepadd}[1]{\textcolor{ShepherdBlue}{#1}}
\else
  \newcommand{\shepdel}[1]{}
  \newcommand{\shepadd}[1]{{#1}}
\fi
\newcommand{\shepreplace}[2]{\shepdel{#1}\shepadd{#2}}

\newcolumntype{C}[1]{>{\centering\arraybackslash}p{#1}}

\algrenewcommand{\algorithmiccomment}[1]{\hfill\textcolor{gray}{// #1}}
\newcommand{\revblue}[1]{#1}
\newcommand{\revorange}[1]{#1}

\newcommand{\revdarkgreen}[1]{#1}
\newcommand{\revbrown}[1]{#1}
\newcommand{\revdarkred}[1]{#1}
\renewcommand{\hl}[1]{#1}
\newif\ifshowcomments

\ifshowcomments
    \newcommand{\zz}[1]{\noindent{\textcolor{purple}{\bf \fbox{ZZ} {\it#1}}}}
\else
    \newcommand{\zz}[1]{}
\fi

\def\BibTeX{{\rm B\kern-.05em{\sc i\kern-.025em b}\kern-.08em
    T\kern-.1667em\lower.7ex\hbox{E}\kern-.125emX}}
\begin{document}



\title{DICE: Enabling Efficient General-Purpose SIMT Execution with Statically Scheduled Coarse-Grained Reconfigurable Arrays}
\author{
\IEEEauthorblockN{Jiayi Wang}
\IEEEauthorblockA{
\textit{University of Washington} \\
Seattle, WA, USA \\
jwang710@uw.edu
}
\and
\IEEEauthorblockN{Ang Da Lu}
\IEEEauthorblockA{
\textit{University of Washington} \\
Seattle, WA, USA \\
angl7@uw.edu
}
\and
\IEEEauthorblockN{Zhichen Zeng}
\IEEEauthorblockA{
\textit{University of Washington} \\
Seattle, WA, USA \\
zczeng@uw.edu
}
\and
\IEEEauthorblockN{Ang Li}
\IEEEauthorblockA{
\textit{University of Washington} \\
Seattle, WA, USA \\
angliz@uw.edu
}
}

\maketitle


\begin{abstract}
While general-purpose graphics processing units (GPGPUs) dominate massively parallel computing through the single-instruction, multiple-thread (SIMT) programming model, their underlying single-instruction, multiple-data (SIMD) execution incurs substantial energy overhead from frequent register file (RF) accesses and complex control logic. 

We present DICE, a novel architecture that addresses these inefficiencies by replacing the SIMD backend with minimal-overhead, statically scheduled coarse-grained reconfigurable arrays (CGRAs) while preserving the well-known SIMT programming model. Unlike SIMD units that execute warps of threads in lockstep, DICE dispatches active threads in a pipelined manner onto the CGRA fabric, where data flows directly between processing elements (PEs), reducing RF accesses for intermediate values. Moreover, DICE organizes threads into groups larger than warps, 
reducing overhead by eliminating the warp-level control logic and further amortizing remaining control overhead across hundreds of threads.

To handle operations with runtime dynamism, such as variable-latency memory loads and data-dependent control flow, while preserving static scheduling, DICE compiles programs into \textit{p-graphs} by partitioning dynamic dependence edges across separate CGRA configurations.
DICE further introduces several key optimizations: double-buffered configuration memory to hide reconfiguration latency, compile-time \textit{p-graph} unrolling to enhance resource utilization, and a temporal memory coalescing unit (TMCU) to merge memory requests from consecutive, pipelined threads. Evaluations on Rodinia benchmarks \shepadd{in Accel-sim} demonstrate that DICE reduces register file accesses by 68\% on average. With equivalent computation and memory resources, DICE's CGRA Processors (CPs) achieve a geometric mean of 1.77--1.90$\times$ dynamic energy efficiency and 42.0\%--45.9\% average power reduction compared to \shepadd{the modeled} NVIDIA Turing Streaming Multiprocessors (SMs), while the full DICE system achieves performance \shepreplace{marginally better than}{comparable to the modeled} Turing \shepreplace{GPUs}{GPU baselines}. DICE demonstrates that spatial pipeline execution can deliver substantial energy savings without sacrificing performance.

\end{abstract}


\section{Introduction}
\label{sec:intro}
The exponential growth in computational demands across domains such as machine learning, scientific computing, and high-performance computing has driven the widespread adoption of massively parallel architectures. Among these, GPGPUs have emerged as the dominant platform for accelerating parallel workloads, extending beyond their original graphics rendering purpose through the SIMT programming model. However, their underlying SIMD execution introduces several inefficiencies that can be optimized. First, GPGPUs communicate intermediate data between instructions through centralized register files (RFs), resulting in frequent RF accesses, which contribute significantly to energy consumption. Second, although threads are grouped into warps to amortize front-end overheads (instruction fetch, decode, and schedule)~\cite{nv_tesla, gpgpu_book}, these control stages remain energy-intensive due to their complexity and frequency of operation. 
Previous studies show that register file accesses and control logic combined consume up to 30\% of total dynamic power~\cite{Hong_Kim_2010, gpuwattch}. \revdarkgreen{More recent studies~\cite{accelwattch, gpu_power_model_2} further indicate that RF and control can account for over 40\% of SM dynamic energy across modern GPU generations, including Pascal~\cite{pascal}, Volta~\cite{volta}, and Turing~\cite{turing}, as well as more recent architectures such as A100~\cite{a100} and H100~\cite{h100}.}

These inefficiencies become increasingly problematic as we approach power density limits in modern process technologies. While specialized units like tensor cores~\cite{volta} address specific workload domains, improving the energy efficiency of general-purpose SIMT execution remains critical for the broad spectrum of GPGPU applications that cannot leverage specialized units.
To address these inefficiencies, we propose \textbf{DICE} (\underline{D}ataflow \underline{I}ntelligent \underline{C}ompute \underline{E}ngine), which replaces the SIMD backend in contemporary GPGPUs with \textbf{minimal-overhead, statically scheduled, spatial-only coarse-grained reconfigurable arrays (CGRAs)} \zz{I don't like minimal-overhead personally, could you replace with another word, e.g., low-cost, area-friendly?}. 

\begin{figure}[t]
  \centering
  \includegraphics[width=\linewidth]{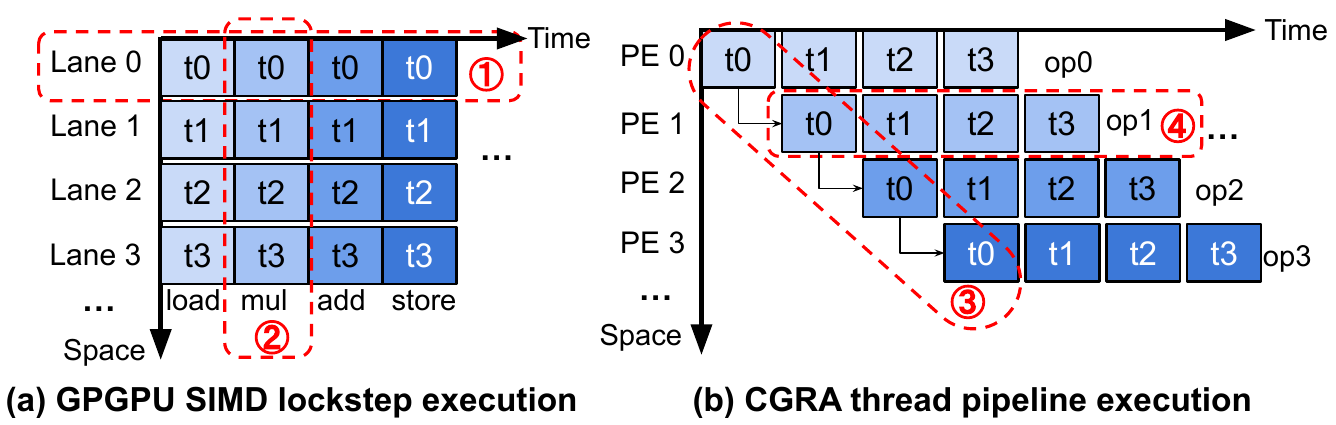}
  \caption{GPGPU SIMD lockstep execution model vs. DICE CGRA thread pipeline execution model. \ding{192}: Each thread is bound to a single lane throughout execution. \ding{193}: All lanes execute the same instruction in lockstep. \ding{194}: Threads flow through different PEs in a pipelined manner. \ding{195}: Each PE is configured to perform a fixed operation for different threads.}

  \label{fig:pipeline_execution}
\end{figure}

Unlike SIMD units that execute instructions in lockstep for a thread warp, DICE executes a massive number of threads on the spatial fabric in a pipelined fashion with an initiation interval ($II$) of one. Fig.~\ref{fig:pipeline_execution} visualizes the differences between GPGPU SIMD lanes and the CGRA spatial pipeline computing model in DICE. On the CGRA fabric, intermediate values are passed directly between PEs via interconnects, greatly reducing RF accesses. DICE further reduces control overhead by using CTAs (Cooperative Thread Arrays) rather than warps as the scheduling granularity, eliminating warp-level control logic overhead and amortizing remaining costs across hundreds of threads. Additionally, the pipeline execution scheme allows DICE to selectively dispatch only active threads to the CGRA, avoiding performance loss from masked SIMD execution under control divergence~\cite{Fung_Sham_Yuan_Aamodt_2007}.

To support operations with runtime dynamism, such as variable-latency memory operations and data-dependent control flows in generic SIMT programs, DICE partitions control-dataflow graphs (CDFGs) into \textit{p-graphs} where the producer-consumer pairs of these operations are strictly separated into different partitions (detailed in Section~\ref{sec:p-graph}). This decoupling ensures that each \textit{p-graph} has a fixed, static schedule on CGRA, eliminating the need for elastic interconnects (e.g., FIFOs) between PEs. DICE uses a lightweight four-stage control pipeline to orchestrate the higher-level \textit{p-graph} execution order (detailed in Section ~\ref{sec:arch}). To hide CGRA reconfiguration latency, DICE employs double-buffered configuration memory for rapid context switching. \zz{will ping-pong buffering better for renaming?}

Beyond the baseline architecture, DICE incorporates two critical optimizations. First, \textit{p-graph} unrolling maps multiple instances of the same \textit{p-graph} onto the CGRA simultaneously, enabling multiple threads to be dispatched per cycle. This optimization is effective for small \textit{p-graphs} that would otherwise underutilize the CGRA fabric. Second, the temporal memory coalescing unit (TMCU) preserves memory efficiency in GPGPUs' memory system despite the pipeline thread execution scheme by buffering and merging requests from consecutive threads that access the same cache line, maintaining coalescing benefits without the lockstep execution requirement.

To summarize, this work makes the following contributions:
\begin{itemize}
\item We identify key inefficiencies in GPGPU SIMT implementation and propose \textbf{DICE}, a novel architecture that replaces the SIMD backend of GPGPUs with statically scheduled CGRAs to address these inefficiencies.
\item We introduce \textit{p-graph} partitioning that enables DICE to handle variable-latency memory operations and control flow without requiring additional hardware mechanisms such as elastic interconnects, dynamic scheduling logic, or flow control between PEs.
\item We examine the bottlenecks in baseline DICE and develop two critical optimizations: \textit{p-graph} unrolling to improve utilization and the temporal memory coalescing unit (TMCU) that preserves memory system efficiency despite the pipeline thread execution scheme.
\item We evaluate DICE by extending Accel-sim~\cite{accelsim} for cycle-accurate performance modeling, and develop a customized \textit{p-graph} compiler. We further develop key components in RTL and synthesize them using Cadence toolchains with FreePDK45~\cite{freepdk45}. 
We build the dynamic power model in AccelWattch~\cite{accelwattch} with Cadence Joules~\cite{joules}.
Evaluation on the Rodinia benchmark suite~\cite{gpu-rodinia} shows that on average DICE reduces RF accesses by 68\% on average. Moreover, DICE's CGRA processors (CPs) achieve a geometric mean of 1.77--1.90$\times$ energy efficiency and 42.0\%--45.9\% average dynamic power reduction relative to \shepadd{the modeled} NVIDIA Turing SM~\cite{turing}, while \shepreplace{the performance of the whole DICE system is marginally better than}{the full DICE system delivers performance comparable to the modeled} Turing \shepreplace{GPUs}{GPU baselines}.
\end{itemize}

\section{Background} 
\label{sec:cgra}

\begin{figure}[t]
  \centering
  \includegraphics[width=0.9\linewidth]{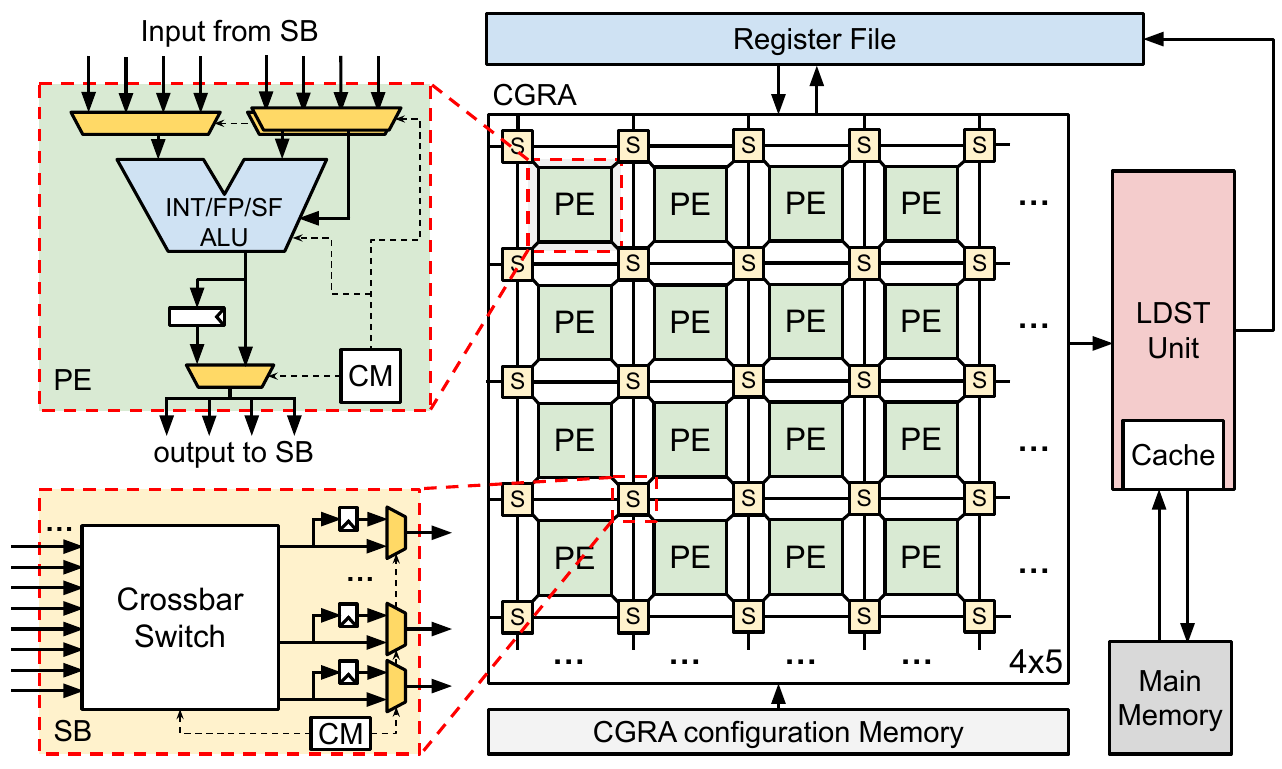}
  \caption{CGRA architecture in DICE machine model. PEs and Switch Boxes (SBs) are scheduled at compile time. The register file serves as the interface between configurations: input operands flow from the register file to the CGRA fabric, while outputs are forwarded to the register file or the load/store Unit (LDST unit) for memory operations.}

  \label{fig:CGRA_model}
\end{figure}

\subsection{SIMT Programming Model} 
The SIMT model in modern GPGPUs offers a simple yet powerful abstraction for parallel programming, allowing developers to write sequential thread code while the hardware manages grouping and scheduling. Unlike SIMD, SIMT supports divergent control flow where threads may follow different branches and allows independent memory accesses without explicit gather or scatter instructions~\cite{nv_tesla, cuda}. GPGPUs implement SIMT through techniques such as warp-based SIMD backend execution, mask-stack for divergence handling, and memory coalescing~\cite{gpgpu_book}. 
DICE preserves the SIMT model to retain these software benefits while introducing a more efficient CGRA-based execution backend.

\subsection{CGRA} 
CGRAs offer unique trade-offs in performance, energy efficiency, and flexibility compared to other hardware architectures such as CPUs, GPUs, ASICs, and FPGAs~\cite{reconfigurable_computing, inefficiency_in_gpc, nre_cost, end_of_moore}.
A CGRA typically consists of an array of PEs operating at word-level granularity, connected via a reconfigurable interconnect. Each PE performs arithmetic or logical operations and forwards data to the next PEs through the interconnect. Configurations for both PEs and interconnects are stored in a dedicated CGRA configuration memory (CM), which is programmed with bitstreams mapping computations onto the spatial fabric.
 
\begin{figure}[tp!]
  \centering
  \includegraphics[width=\linewidth]{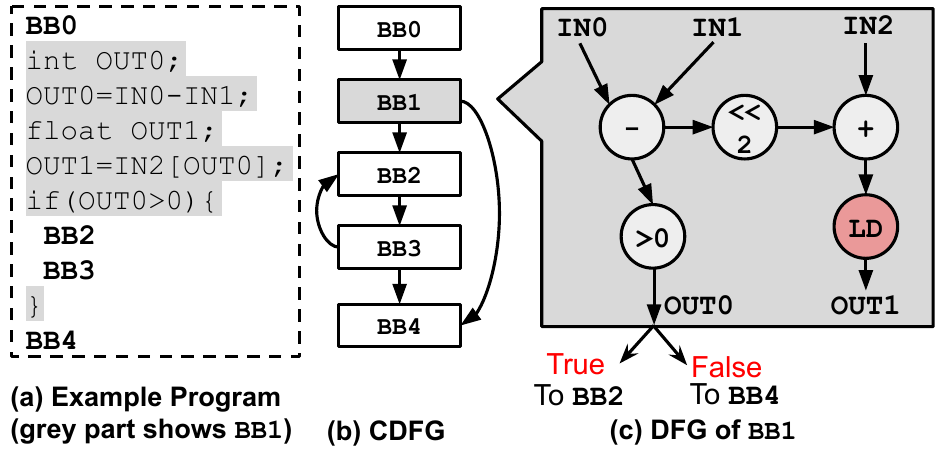}
  \caption{Control-data flow graph example (LD: Load, BB: Basic block)}

  \label{fig:CDFG}
\end{figure}

Program mapping depends on CGRA capabilities. In basic CGRAs without control-flow support, the program’s control-dataflow graph (CDFG) is partitioned into basic blocks (BBs), each compiled into a dataflow graph (DFG) of pure data dependencies (control handled externally). Fig.~\ref{fig:CDFG} shows an example CDFG and a corresponding DFG for a simple program. More advanced CGRAs support direct mapping of control flow onto the fabric through mechanisms such as $\phi$-nodes (select operations) for merging values from different paths~\cite{hycube}, predicated execution~\cite{AHA}, or dedicated control flow operations~\cite{riptide}. These approaches enable mapping larger CDFG portions, potentially the entire program, as a single configuration, improving performance by reducing reconfiguration overhead and enabling aggressive optimizations at the cost of additional hardware complexity and lower energy efficiency.

The CGRA design space is broad~\cite{design_space, cgra_survey_1, cgra_survey_2}, but two dimensions are particularly relevant to DICE:

The first dimension distinguishes \textbf{spatial-only} CGRAs from \textbf{temporal-spatial} CGRAs: Spatial-only CGRAs~\cite{amber, snafu, riptide} keep a single configuration active throughout execution, with each PE fixed to one operation. Temporal–spatial CGRAs~\cite{hycube, opencgra, cgra_me, vecpac, plasticine} store multiple configurations and can switch between them every cycle, time-multiplexing resources to improve flexibility and utilization at the cost of extra silicon area and switching power.

The second dimension distinguishes \textbf{statically scheduled} CGRAs from \textbf{dynamically scheduled} CGRAs: Statically scheduled CGRAs~\cite{hycube, opencgra, cgra_me, vecpac, amber} execute operations in a fixed order determined at compile time. Dynamically scheduled CGRAs fire operations when operands are ready, using flow-control (valid-ready or credit-based)\cite{plasticine, snafu, riptide} or tagged-token dataflow~\cite{sgmf,vgiw,dmt-cgra,tag_token_1,tag_token_2} to distinguish multiple in-flight instances. Dynamic scheduling tolerates latency variability and irregular control/data patterns, but requires extra hardware for tagging, matching, and buffering.

\revblue{A recent study~\cite{pe_eval} quantifies the area overhead of other CGRA styles relative to a statically scheduled, spatial-only CGRA. Temporal-spatial CGRAs incur a 1.6$\times$ area overhead (+160\%), while dynamically scheduled CGRAs add 1.1$\times$ overhead (+110\%) with flow-control and 4.8$\times$ overhead (+480\%) with tagged-token dataflow.\footnote{The statically scheduled, spatial-only CGRA baseline in~\cite{pe_eval} is denoted as ``Systolic,'' while temporal-spatial CGRAs are denoted as ``CGRA.''}}

\section{\textit{p-graph} and DICE Machine Model} 
\label{sec:p-graph}

To minimize hardware overhead, \textbf{DICE employs the statically scheduled, spatial-only CGRAs} as illustrated in Fig.~\ref{fig:CGRA_model}. PEs can support integer (INT), floating-point (FP), or special functional (SF) operations according to the arithmetic logic units (ALU) types, with a 1-bit control signal for predicated execution. The interconnect uses statically scheduled, wire-switched routing similar to the one in AHA~\cite{AHA}.

\subsection{\textit{p-graph} formation}
To support operations with runtime dynamism, including memory operations with unpredictable, variable latency, and data-dependent control flow operations, DICE partitions the CDFG into multiple subgraphs that are later compiled into different CGRA configuration bitstreams, separating producer-consumer pairs of these operations across different CGRA configurations and moving such dependence edges to CGRA boundaries for external handling. We call these partitioned subgraphs \textbf{\textit{p-graphs}}. As shown in Fig.~\ref{fig:dfg_partition}, each \textit{p-graph} satisfies the following constraints:

\begin{figure}[tp!]
    \centering
    \includegraphics[width=\linewidth]{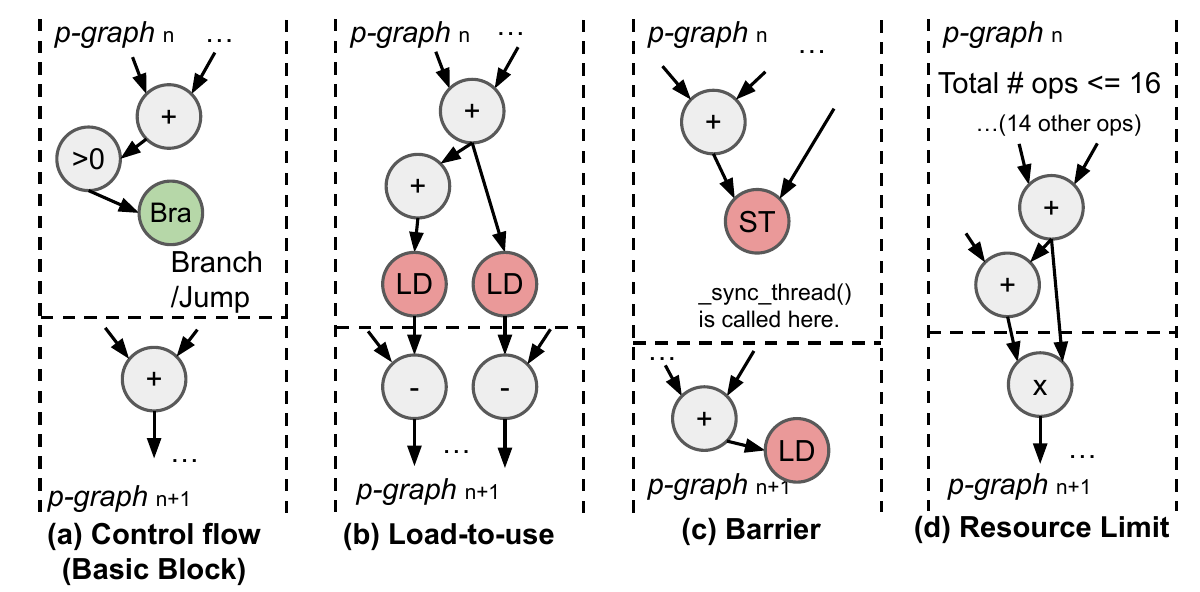}
    \caption{\textit{p-graphs} constraints}

    \label{fig:dfg_partition}
\end{figure}

\begin{enumerate}
\item \textbf{Control flow constraint (Fig.~\ref{fig:dfg_partition}(a)):} No control divergence (jump or branch instructions) within a \textit{p-graph} (Same constraint in basic block). This constraint can be relaxed with predicated execution (Section~\ref{sec: predicated execution}).
\item \textbf{Memory load constraint (Fig.~\ref{fig:dfg_partition}(b)):} No memory load-to-use dependencies within a \textit{p-graph}.
\item \textbf{Barrier constraint (Fig.~\ref{fig:dfg_partition}(c)):} Synchronization barriers terminate a \textit{p-graph}.
\item \textbf{Resource constraint (Fig.~\ref{fig:dfg_partition}(d)):} resource usage must fit within CGRA capacity.
\end{enumerate}

\begin{table*}[t]
\centering
    \caption{\textit{p-graph} Metadata Components}
    \label{tab:metadata}
    \small
    \begin{tabular}{ccl}
        \toprule
        Field & Value Type &Description \\
        \midrule
        BITSTREAM\_ADDR & 32-bit address & Address of the corresponding CGRA configuration bitstream \\
        BITSTREAM\_LENGTH & 8-bit value & Bitstream size in bytes \\
        UNROLLING\_FACTOR & 2-bit value & Max thread unrolling factor (see Section~\ref{sec:unroll}) \\
        LAT & 8-bit value & CGRA fabric latency for the \textit{p-graph} \\
        IN\_REGS & 34-bit bitmap &Input registers bitmap of the \textit{p-graph} \\

        OUT\_REGS & 34-bit bitmap &Direct output registers bitmap from CGRA fabric of the \textit{p-graph} \\
        LD\_DEST\_REGS & $4\times6$-bit value & Destination register indexes for memory loads (max request ports $\times$ index width)  \\
        NUM\_STORES & 3-bit value & Number of stores per thread of this \textit{p-graph} \\
        BRANCH\_* & 32-bit value & Branch/Jump metadata associated with this \textit{p-graph} \\
        BARRIER & 1-bit bool & Barrier indicator, all previous blocks must retire before executing this \textit{p-graph} \\
        PARAMETER\_LOAD & 1-bit bool & Special \textit{p-graph} for loading kernel parameters (see Section~\ref{sec:arch}) \\
    \bottomrule
\end{tabular}
\end{table*}

These constraints ensure that each \textit{p-graph} is statically analyzable, has fixed latency, and is mappable within the spatial constraints of the DICE CGRA fabric. 

\subsection{DICE Machine Model}
\begin{figure}[t]
    \centering
    \includegraphics[width=\linewidth]{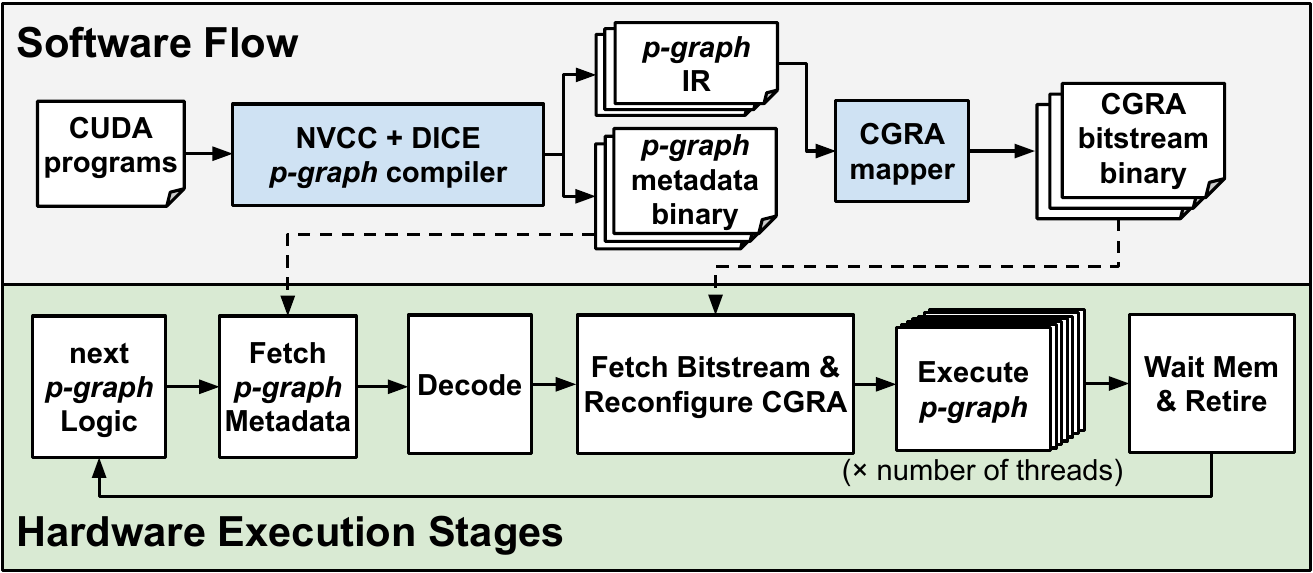}
    \caption{DICE software flow and hardware execution stages} 
    \label{fig:execution_model}
\end{figure}

\begin{figure}[t]
    \centering
    \includegraphics[width=\linewidth]{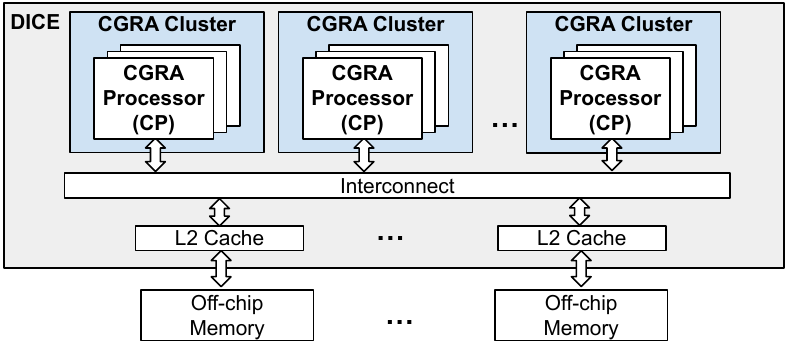}
    \caption{DICE top-level organization. CGRA Processors (CPs) are grouped into clusters, which connect to the submemory system through a shared interconnect.}

    \label{fig:top}
\end{figure}

DICE employs the register file for inter-\textit{p-graph} data communication. To handle memory operations at the boundary of each \textit{p-graph}, DICE employs a dedicated load/store unit (LDST Unit) placed alongside the CGRA as illustrated in Fig.~\ref{fig:CGRA_model}. Operands are read from the register file into the CGRA, and CGRA outputs are either directly written back for subsequent \textit{p-graphs} or routed to the LDST Unit for memory accesses. The LDST Unit fetches data from the memory system into the register file, completing the execution loop.

Fig.~\ref{fig:execution_model} illustrates both the software flow and hardware execution model of DICE across multiple \textit{p-graphs}.

\textbf{Software flow:} Starting from programs written in CUDA~\cite{cuda}, NVCC~\cite{nvcc} compiles the high-level code into PTX~\cite{ptxisa88}, an assembly-like intermediate representation (IR). The DICE compiler then: (1) constructs CDFGs from PTX IR, (2) partitions them into \textit{p-graphs}, and (3) generates \textit{p-graph} IR with associated metadata. Table~\ref{tab:metadata} summarizes the metadata fields used to describe each \textit{p-graph}. Like in conventional CPU/GPGPU instruction set architectures (ISAs), inputs and outputs are encoded as register numbers. The \textit{p-graph} IR is then passed to a CGRA mapper, which performs placement and routing on the CGRA fabric and generates the final CGRA configuration bitstream.

\textbf{Hardware execution stages:} The execution stages across \textit{p-graphs} resemble traditional von Neumann-style control pipelines with instructions. As illustrated in Fig.~\ref{fig:execution_model}, the execution flow begins with (1) the next \textit{p-graph} logic selecting the subsequent configuration. (2) The corresponding metadata is fetched and (3) decoded to identify the bitstream address and configure CGRA peripherals. (4) This CGRA bitstream is subsequently retrieved and loaded into the CGRA configuration memory. (5) Once reconfiguration completes, threads are dispatched to the CGRA to execute the current \textit{p-graph} in a pipelined fashion with $II=1$ as shown in Fig.~\ref{fig:pipeline_execution}. (6) Following CGRA computation completion, the system waits for outstanding memory operations before retiring the current \textit{p-graph} and proceeding to the next. This cycle continues until all kernel \textit{p-graphs} have executed on all threads.

\begin{figure*}[t]
    \centering
    \includegraphics[width=\linewidth]{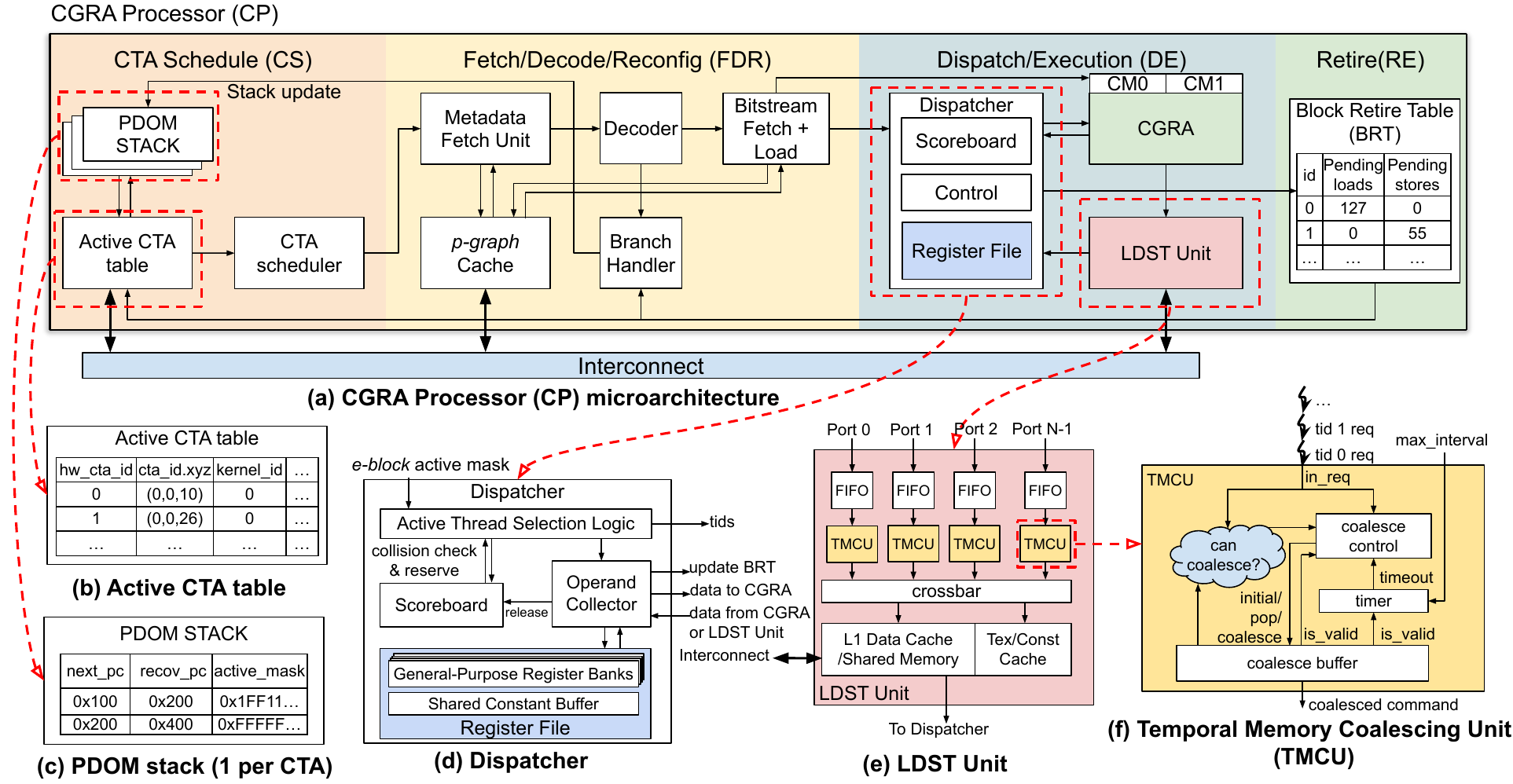}
    \caption{CGRA Processor (CP) microarchitecture }

    \label{fig:cgra_core}
\end{figure*}

\section{DICE Architecture}
\label{sec:arch}
As shown in Fig.~\ref{fig:top}, DICE follows a hierarchical organization similar to contemporary GPGPUs~\cite{gpgpu_book}: compute resources are grouped into CGRA Clusters (CC), each containing multiple CGRA Processors (CPs), with clusters connected to main memory via the system interconnect.

When a kernel launches, each CTA is assigned to a CP. Unlike GPGPUs that schedule at warp granularity, DICE operates at the CTA level: when the CGRA is reconfigured for a new \textit{p-graph}, all threads from the entire CTA are dispatched sequentially to the CGRA in a pipeline manner. This approach eliminates the need for warp-level control logic and amortizes the overhead of \textit{p-graph} metadata/bitstream fetching and CGRA reconfiguration across hundreds of threads, further improving energy efficiency. At runtime, when a CTA is scheduled to execute a specific \textit{p-graph}, DICE instantiates a corresponding execution block, referred to as an \textit{e-block}. An \textit{e-block} is a dynamic runtime entity that encapsulates the execution of a \textit{p-graph} for currently active threads in a given CTA. Conceptually, \textit{p-graphs} and \textit{e-blocks} are analogous to static and dynamic instructions, respectively, in conventional CPU/GPGPU terminologies.

The key design goal of the CP is to maximize CGRA utilization through efficient pipelining and organization.

\subsection{CGRA Processor (CP) Baseline Microarchitecture}
\begin{figure*}[t!]
    \centering
    \includegraphics[width=\linewidth]{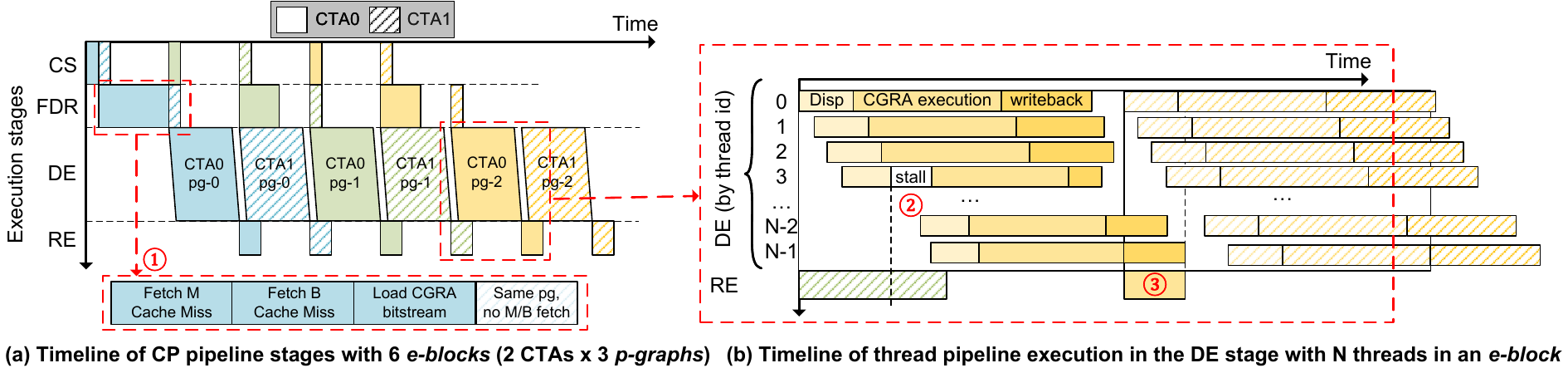}
    \caption{CGRA Processor (CP) pipeline execution example (M: metadata, B: CGRA bitstream, pg: \textit{p-graph}, Disp: dispatch). \ding{192}: The CTA scheduler selects \textbf{CTA1} after \textbf{CTA0} as the next \textit{e-block} to run due to the same \textit{p-graph} as the previous one. The FDR stage of \textit{pg-0} for \textbf{CTA1} is much shorter since metadata and bitstream from the previous fetch can be reused. \ding{193}: Dispatch stalls when scoreboard detects data dependency on pending memory writeback.  \ding{194}: As soon as all threads complete CGRA execution, the \textit{e-block} enters the RE stage and retires once all memory operations finish.}
    \label{fig:timeline}
\end{figure*}

Fig.~\ref{fig:cgra_core}(a) illustrates the four pipeline stages of a CP in DICE: CTA schedule (CS), Fetch/Decode/Reconfig (FDR), Dispatch/Execution (DE), and Retire (RE). Fig.~\ref{fig:timeline} presents an example execution timeline involving 2 CTAs assigned to a single CP, where the kernel is partitioned into 3 \textit{p-graphs}. Blocks in the same color and pattern represent the same \textit{e-block} passing through different stages.

\subsubsection{CTA Schedule (CS)} This stage selects the next CTA and forms an \textit{e-block} for the next pipeline stages. The \textbf{Active CTA Table} (Fig. ~\ref{fig:cgra_core}(b)) stores CTAs assigned by the kernel driver, along with their associated kernel metadata. Each active CTA is linked with a dedicated \textbf{Post-Dominator (PDOM) Stack} (Fig.~\ref{fig:cgra_core}(c)) similar to ones used in NVIDIA Fermi GPGPUs~\cite{gpgpu_book, Fung_Sham_Yuan_Aamodt_2007} to track control divergence. The PDOM stack can handle both nested control flow and skipped computation entirely by stacking the next program counter (PC), reconvergence PC, and the active thread mask of each divergent path.

\textit{Scheduling Strategy:} To maximize metadata and bitstream reuse to reduce overhead, the \textbf{CTA Scheduler} prioritizes CTAs whose next PC matches that of recently dispatched \textit{e-blocks}, while applying a round-robin policy among otherwise active CTAs. Once a CTA is selected, a new \textit{e-block} is instantiated with the CTA's metadata and the next \textit{p-graph} metadata PC and then forwarded to the next stage.

\subsubsection{Fetch/Decode/Reconfig (FDR)} \label{sec:fdr} In this stage, the \textbf{Metadata Fetch Unit} fetches the metadata for the corresponding \textit{p-graph} from \textbf{\textit{p-graph} Cache} (analogous to GPGPU's instruction cache), which is then decoded by the \textbf{Decoder}. The \textbf{Branch Handler} processes control flow information embedded in the \textit{p-graph} metadata and determines whether the next \textit{p-graph} of this CTA can be speculatively identified before this \textit{e-block} completes execution. Then it updates the PDOM stack of the corresponding CTA accordingly.

\textit{Branch Prediction Strategy}: The branch handler employs a backward-taken forward-not-taken (BTFNT) prediction scheme. In Fig.~\ref{fig:CDFG}, both \textbf{BB1} and \textbf{BB3} predict \textbf{BB2} as their successor. This predictor requires minimal hardware yet performs well when both paths contain active threads, allowing the CS and FDR stages to proceed without stalls. However, the \textit{e-block} must wait before entering the DE stage until thread activeness is resolved; mispredicted \textit{e-blocks} are discarded.

\textit{Handling Synchronization Barriers}: When metadata indicates a synchronization barrier, the FDR stage must ensure all prior operations complete before proceeding to the DE stage. Unlike GPGPUs that track individual warp arrivals at barriers, DICE's CTA-level execution eliminates this complexity. The FDR stage simply waits for signals from the RE stage confirming that all prior \textit{e-blocks} from the CTA have completed memory operations.

\textit{Double-buffered CGRA Configuration Memory}: After decoding the address of the bitstream, the \textbf{Bitstream Fetch and Load} unit checks whether the required CGRA configuration bitstream is already resident in either one of two configuration memories (CM0/CM1, shown in Fig.~\ref{fig:cgra_core}(a)). Missing bitstreams are fetched from the \textit{p-graph} cache and loaded into the inactive configuration memory, enabling reconfiguration without interrupting ongoing CGRA execution.

An \textit{e-block} is considered ready to proceed to the next stage once all the following conditions are met: metadata decoding is complete, the CGRA bitstream is loaded, active thread information is resolved, and any barrier requirements (all prior \textit{e-blocks} of the CTA completed and retired) are satisfied.

\subsubsection{Dispatch/Execution (DE)}
This stage executes threads on the CGRA fabric through three main components: the \textbf{Dispatcher}, the \textbf{CGRA}, and the \textbf{LDST Unit}.  

\textit{Selective Dispatch}: The Dispatcher (Fig.~\ref{fig:cgra_core}(d)) orchestrates thread sequencing, manages operands, and tracks memory requests through four subcomponents: \textbf{Active Thread Selection Logic}, \textbf{Scoreboard}, \textbf{Register File}, and the \textbf{Operand Collector}. The Active Thread Selection Logic extracts the sequence of active threads from the \textit{e-block} active mask and outputs them in ascending thread ID (tid) order. In this process, DICE skips inactive threads entirely, avoiding performance loss in masked SIMD execution.

\textit{Conflict-free Register File and Management}:
\label{sec:dice rf} 
The register file (RF) stores intermediate values exchanged across \textit{e-blocks}. It includes multiple \textbf{General-Purpose Register Banks} storing private registers of each thread, and a \textbf{Shared Constant Buffer} for constant values like function arguments that are shared across all threads in the same kernel. To maintain $II=1$, the RF employs a banking scheme where registers with different indices map to separate banks, while same-index registers across threads share a bank. This organization enables conflict-free access to all of a thread's registers simultaneously, while avoiding energy waste from fetching unnecessary registers with fewer banks. The Operand Collector coordinates data movement—fetching operands, managing writeback, and tracking memory operations—while the Scoreboard prevents data hazards by verifying operand availability before actually dispatching a thread.

\label{sec: predicated execution}
\textit{Predicated Execution and Writeback}: CGRA outputs are routed to both the RF (via the Operand Collector) and the \textbf{LDST Unit} (Fig. ~\ref{fig:cgra_core}(e)). The LDST Unit in baseline DICE resembles its GPGPU counterpart with a fixed number of ports, except memory requests from different threads arrive sequentially through the same port over pipeline cycles, buffered by a FIFO. (The optimized DICE design incorporates a temporal memory coalescing unit (TMCU), described in Section~\ref{sec:tmcu}.) As described in Section~\ref{sec:p-graph}, each PE has a 1-bit control signal input for predicated execution. These control signals also serve as CGRA output controls: they enable/disable register writeback and mark memory requests as valid/invalid. This predication support allows DICE to translate control flow into predicated execution, enabling the merging of \textit{p-graphs} that were originally separated due to control divergence. Instead of being separated into different \textit{p-graphs} for divergent paths, both paths execute within a single \textit{p-graph} with their operations selectively enabled by predication bits, improving utilization and reducing reconfiguration overhead. 

\label{sec:stall}
\textit{Stall Conditions}: The dispatcher stalls when: (1) the scoreboard detects the current dispatching thread's input registers are still awaiting memory loads, (2) any FIFO in the LDST Unit runs out of credit (reaches the capacity), or (3) no available threads to dispatch (Idle state). The idle state can occur when no ready \textit{e-block} is available in the previous stage, or when all threads of the current \textit{e-block} have been dispatched, but CGRA computation is still in progress.

\shepadd{\textit{Pipeline Fill/Drain Overhead}: The CGRA incurs bubbles during pipeline fill and drain, adding an extra latency of $p$ cycles to the execution of a \textit{p-graph} with latency $p$. For a CTA with $t$ threads, the total execution time is therefore $t + p$ cycles when dispatch is not stalled. Since each CTA typically contains hundreds of threads, the relative fill/drain bubble overhead of $p/t$ on overall performance is small.}

\subsubsection{Retire (RE)}
To avoid occupying CGRA compute resources after all threads in an \textit{e-block} have completed CGRA execution but are still waiting on memory responses, DICE employs the RE stage with the \textbf{Block Retire Table (BRT)}. The BRT, illustrated in Fig.~\ref{fig:cgra_core}, can hold multiple \textit{e-blocks} and tracks the status of their outstanding memory operations. An \textit{e-block} transitions from the DE stage to the RE stage once all of its threads have been dispatched and have completed computation operations on the CGRA fabric, as shown in Fig.~\ref{fig:timeline}(b)\ding{194}. While in the BRT, \textit{e-blocks} are monitored for completion of pending memory requests. Once all outstanding memory operations have completed for an \textit{e-block}, the BRT retires the \textit{e-block} by removing it from the BRT and updates the remaining \textit{e-block} status to Branch Handler and Active CTA table. The RE stage allows the CGRA fabric to continue executing subsequent \textit{e-blocks} without stalling on memory latency, thereby improving throughput and overall resource utilization.

\subsection{Optimizations upon Baseline DICE}

\begin{algorithm}[tp!]
\caption{Temporal Memory Coalescing Logic}
\label{alg:tmcu}
\small
\begin{algorithmic}[0]  
\Function{OutputCommand}{in\_req} \Comment{Run every cycle}
  \If{coalesce\_buffer.is\_valid()}
    \State timer.decrement()
  \EndIf
  \If{timer.timeout()}
    \State coalesce\_buffer.pop() \Comment{Pop the req in buffer}
    \State timer.reset(max\_interval)
  \EndIf
  \If{in\_req.is\_valid()} \Comment{Process new request}
    \If{not coalesce\_buffer.is\_valid()}
      \State coalesce\_buffer.initial(in\_req)
    \ElsIf{coalesce\_buffer.can\_coalesce(in\_req)}
      \State coalesce\_buffer.coalesce(in\_req)
    \Else
      \State coalesce\_buffer.pop() \Comment{Pop to cache}
      \State timer.reset(max\_interval)
      \State coalesce\_buffer.initial(in\_req)
    \EndIf
  \EndIf
\EndFunction
\end{algorithmic}
\end{algorithm}

\subsubsection{Thread Unrolling for Enhanced Utilization}
\label{sec:unroll}
While predicated execution enables the merging of divergent paths into larger \textit{p-graphs}, some remain inherently small due to other constraints, leading to CGRA underutilization. DICE addresses this through thread unrolling similar to loop unrolling in DySER~\cite{dyser}.

Key challenges arise when unrolling threads for CGRA execution: (1) maintaining $II=1$ with multiple simultaneous threads, since unrolling by 2$\times$ with $II=2$ provides no benefit over single-thread dispatch with $II=1$; (2) avoiding complex runtime hardware mechanisms and corresponding metadata; and (3) keeping compiler complexity manageable while enabling optimization flexibility.

DICE employs hardware/software co-design to address these challenges. On the hardware side, DICE adopts a hardware swizzled register bank mapping ~\cite{swizzle}, where each thread's private registers are shifted by one bank relative to neighboring threads. Assuming the total number of logical registers is $N_r$, the number of physical banks is also equal to $N_r$ as stated in section~\ref{sec:dice rf}. Physical thread $T$'s register $R_r^T$ in a CP is mapped to bank ($(r+T)$ mod $N_r$), ensuring adjacent threads access different banks. The dispatcher supports multiple lanes but restricts co-dispatched threads to fixed intervals of $K$; that is, only threads $(T,T+K,T+2K,… )$ can be dispatched together. The maximum simultaneous threads that can be dispatched is $N_{Tmax}$. 

The DICE compiler is aware of the hardware's swizzling policy and uses this information to statically determine the maximum safe unrolling factor for each individual \textit{p-graph}. Given a set of input registers $(R_{a_1},R_{a_2},..R_{a_{p-1}})$ for a \textit{p-graph}, the compiler finds the maximum unrolling factor that avoids bank conflicts and respects hardware resource limits.

When bank conflicts limit unrolling, the compiler is able to adjust register allocation at compile time to get higher unrolling factors. In the DICE model used for evaluation, $N_r = 32$, $N_{Tmax}=4$, with $K=8$ for 4× unrolling and $K=16$ for 2× unrolling (3× unsupported).

\sethlcolor{yellow}
\begin{table*}[t]
\centering
\caption{DICE and NVIDIA Turing GPGPU (RTX2060S) configurations in evaluation.}
\label{tab:system config}
\footnotesize
\begin{tabular}{l|l|C{4.8cm}|C{4.8cm}}
\toprule
\textbf{Category} & \textbf{Component} & \textbf{RTX2060S~\cite{turing}} & \textbf{DICE (This work)} \\
\midrule
\multirow{3}{*}{\textbf{Organization}} 
 & Compute Clusters & 34 SMs & 34 CGRA Clusters (CC) \\
 & Core Partition & 4 subcores/SM & 4 CPs/Cluster \\
 & Max threads & 1024/SM & 2048\textsuperscript{\#}/Cluster \\
\midrule
\multirow{3}{*}{\textbf{Functional Units}} 
 & \multirow{2}{*}{ALUs} & 16 CUDA Cores / subcore\textsuperscript{*} & 16 PEs\textsuperscript{†} / CP \\
  &  & 2176 CUDA Cores total & 2176 PEs total \\
\midrule
\multirow{4}{*}{\textbf{Memory}} 
 &  \multirow{2}{*}{Register File} & 256 KB/SM & \hl{32 Banks/CP}, 256 KB/Clusters \\
 & &8.5 MB total &  8.5 MB total\\
 & L1 Cache / Shared Memory & 96 KB/SM, 3.2 MB total & 96 KB/Cluster, 3.2 MB total \\
 & L2 Cache & \multicolumn{2}{c}{4096 KB total (16-way, 64 sets)} \\
 & DRAM & \multicolumn{2}{c}{GDDR6 DRAM, 8 channels} \\
\midrule
\multirow{2}{*}{\textbf{Frequency}} 
 & Core, Interconnect, L2 Cache & \multicolumn{2}{c}{1470 MHz} \\
 & DRAM & \multicolumn{2}{c}{1750 MHz} \\
\bottomrule
\end{tabular}
\begin{FlushLeft}
  \footnotesize
  \textsuperscript{\#}DICE allocates 32 private 32-registers per thread. With the same total RF capacity, DICE can keep more thread contexts resident because spatial execution reduces register pressure.
  \textsuperscript{*}In RTX2060S, each CUDA core has separate INT and FP units that can execute simultaneously. 
  \textsuperscript{†}In DICE, each PE can perform either one INT or FP operation per cycle, but not both concurrently.
  \end{FlushLeft}
\end{table*}

\subsubsection{Temporal Memory Coalescing Unit}
\label{sec:tmcu}
Memory coalescing 
reduces pressure on the memory system by combining multiple narrow memory requests into fewer, wider memory transactions, improving memory bandwidth utilization, lowering access latency, and reducing energy consumption. In GPGPUs, threads within a warp issue memory requests simultaneously, enabling the coalescing unit to identify and merge accesses in parallel
. GPGPU programs are typically optimized to exploit this coalescing behavior.

However, this coalescing strategy does not directly apply to DICE,  
where memory requests arrive sequentially over multiple cycles through a shared port, lacking the simultaneous visibility that enables GPGPU-style coalescing. 

\revdarkred{While Miss Status Holding Registers (MSHRs)~\cite{mshr} track outstanding misses to support non-blocking caches, they are not a coalescing mechanism: they retain per-miss state after a request is issued, but do not opportunistically merge neighboring requests into a wider transaction.}

To preserve coalescing benefits in pipelined execution, we introduce the Temporal Memory Coalescing Unit (TMCU). As shown in Fig.~\ref{fig:cgra_core}(f) and Algorithm~\ref{alg:tmcu}, the TMCU maintains a coalescing buffer for gathering and merging compatible memory requests. When a new memory request arrives, the TMCU evaluates whether it can be coalesced with the buffered transaction based on request type and address alignment (can\_coalesce() function in Algorithm~\ref{alg:tmcu}). If the request can be coalesced, it is merged into the command in the buffer; otherwise, the existing buffered command is popped and sent to L1 caches, after which the new request becomes the base command in the coalesce buffer.

To prevent commands from remaining in the coalescing buffer indefinitely due to a lack of subsequent requests, the TMCU includes a timeout mechanism. A timer tracks how long a coalesced command has remained in the buffer, and once it exceeds a preset max interval, the command is flushed to the memory system to ensure forward progress. Under typical access patterns where consecutive threads access contiguous memory, the TMCU achieves coalescing equivalent to traditional GPGPU coalescing units. In rare corner cases (e.g., when requests from threads 0 and 2 are able to coalesce but thread 1 is not), the TMCU may fail to merge them due to its in-order and temporal processing constraints. However, such scenarios are infrequent in well-optimized GPGPU programs.

\revorange{
\subsection{DICE Scalability}
\label{sec:scale}
DICE exhibits two forms of scalability, similar to other spatial architectures~\cite{scalesim-v3}: \textit{scale-up} and \textit{scale-out}.

\textbf{Scale-up (Larger CGRAs).}
Increasing the CGRA size enables larger \textit{p-graphs} to be mapped onto the fabric, allowing more operations to execute spatially and improving performance. A larger fabric can also eliminate certain resource-induced partitioning boundaries, reducing intermediate register file storage previously required due to capacity constraints. Furthermore, \textit{p-graphs} that are not resource-bound can exploit additional spatial capacity through further unrolling to maintain high utilization. However, for small \textit{p-graphs} that have already reached the maximum supported unrolling factor, increasing CGRA size alone may lead to lower PE utilization.

\textbf{Scale-out (More CPs).}
Scale-out increases the number of CGRA clusters while keeping each CP microarchitecture unchanged. This approach mirrors how GPU families scale by varying the number of SM per die (e.g., RTX2060S (34 SMs), Quadro RTX5000 (48 SMs), and Quadro RTX6000 (72 SMs) in NVIDIA Turing GPGPUs family~\cite{turing}) while preserving the SM design. Since each CP is unchanged, scale-out increases peak compute capacity and typically improves throughput in the same way that adding more SMs does in GPUs, subject to workload parallelism and memory-system limits.
}

\begin{table}[t]
\centering
\caption{Evaluated kernels and launch configurations.}
\label{tab:kernels}
\begin{tabular}{l|c|c|c|c}
\toprule
\textbf{Benchmark: Kernel} & \textbf{\#\textit{p-g}} & \textbf{B} & \textbf{G} & \textbf{T} \\
\midrule
NN: euclid              & 4  & 256 & 2048 & 524288 \\

BFS-1: kernel           & 10 & 512 & 128  & 65536 \\
BFS-2: kernel2          & 4  & 512 & 128  & 65536 \\

BPNN-1: layerforward    & 10 & 256 & 256  & 65536 \\
BPNN-2: adjust\_weights & 7  & 256 & 256  & 65536 \\

SC: compute\_cost       & 12 & 512 & 128  & 65536 \\

GE-1: Fan1              & 5  & 512 & 1    & 512 \\
GE-2: Fan2              & 6  & 256 & 169  & 43264 \\

HS: calculate\_temp     & 13 & 256 & 1849 & 473344 \\

PF: dynproc\_kernel     & 8  & 256 & 544  & 139264 \\
\bottomrule
\end{tabular}
\begin{tablenotes}
      \item \#\textit{p-g}: Number of \textit{p-graphs}; B: Block/CTA size (flattened); G: Block grid size (flattened); T: Total threads ($T = B \times G$).
    \end{tablenotes}
\end{table}


\section{Experimental Methodology}
\subsection{Basic Power, Performance, Area Evaluation Methodology}
\label{sec:methodology}
\textbf{Hardware modeling}:
We model DICE by extending the Accel-sim~\cite{accelsim} cycle-accurate GPGPU simulator. We further develop RTL for DICE-specific components and synthesize them using the open-source FreePDK45 process design kit~\cite{freepdk45} with the Cadence toolchain. We also use CACTI~\cite{cacti} to accurately model memory area and timing characteristics, and all results are subsequently scaled to 12nm FinFET technology node with scaling factors in ~\cite{scale} to compare with the RTX2060S baseline. 

To estimate the dynamic power and energy consumption of DICE, we 
extend the AccelWattch~\cite{accelwattch} power modeling framework to construct the dynamic power model of DICE. 
\revbrown{To ensure that the reported energy differences arise from architectural changes rather than circuit-level modeling discrepancies, we apply a consistent methodology to both designs.
We use AccelWattch's per-access energy for hardware components that are the same across DICE and the baseline GPU (e.g., L1 cache, Shared Memory, Compute).
SRAM blocks introduced by DICE are modeled with CACTI. As for DICE-specific control logic, we estimate energy consumption with RTL implementation and realistic activity trace-based analyses using Cadence Joules~\cite{joules}.
}

\textbf{Compilers}:
We develop a customized DICE \textit{p-graph} compiler, which translates PTX~\cite{ptxisa88} code compiled from NVCC~\cite{nvcc} to CDFGs and partitions CDFGs into \textit{p-graphs}. The compiler incorporates an optimization pass to generate performance-optimized \textit{p-graphs}. While our compiler handles most kernels automatically, we apply manual optimizations in some cases to ensure architecture evaluation is not limited by compiler maturity.

\textbf{Baseline}:
\label{sec:baseline}
We compare DICE against the NVIDIA Turing GPGPU RTX2060 Super~\cite{turing} (RTX2060S) in terms of overall performance and dynamic power and energy consumption at the core level (DICE CPs vs. RTX2060S SMs). To our knowledge, the RTX2060S is currently the most advanced model with detailed dynamic power modeling support in AccelWattch~\cite{accelwattch}. To ensure a fair comparison, DICE is configured to match the RTX2060S in both compute resource count and total cache/SRAM capacity as shown in Table~\ref{tab:system config}. DICE CPs use 4$\times$5 CGRAs composed of 16 PEs and 4 Special Function Units (SFUs). 
In addition to this cross-architecture comparison, we evaluate internal DICE baselines in which thread unrolling and memory coalescing are selectively disabled or enabled in isolation, allowing us to quantify the individual and combined contributions of these techniques to performance. Specifically:
\begin{itemize}
\item \textbf{\revorange{DICE-na\"ive}}: Both \textit{p-graph} unrolling and TMCU are off. 
\item \textbf{\revorange{DICE-na\"ive+unroll}}: Only unrolling is enabled while TMCUs are disabled. DICE supports unrolling factors equal to 2 and 4. 
\item \textbf{\revorange{DICE-na\"ive+TMCU}}: Only TMCUs are enabled with \textit{max\_interval} equals 8, which matches the 32-byte L1 data cache sector size with 4-byte data access. Unrolling is disabled.
\item \textbf{\revorange{DICE}}: Both optimizations are applied.
\end{itemize}

\textbf{Benchmarks}:
We evaluate DICE using diverse benchmarks from the Rodinia benchmark suite~\cite{gpu-rodinia}.
Table~\ref{tab:kernels} summarizes the selected benchmarks and their kernel configurations.

\revorange{
\subsection{Scalability Evaluation Methodology}
\label{sec:scale_eval}

\subsubsection{Scale-up (Larger CGRAs)}
To study CGRA scale-up, we compare the baseline 16-PE DICE with DICE-U, a 32-PE design that doubles the CGRA size per CP. We scale RF, L1, and supported threads per CP proportionally. To keep per-cluster compute and memory constant, we halve the number of CPs per cluster, isolating the performance–RF-access trade-off of larger CGRAs (Table~\ref{tab:dice_40pe}).

\subsubsection{Scale-out (More CPs)}
To evaluate system-level scaling, we model larger Turing-class GPUs in Accel-sim by increasing the number of SMs to match the Quadro RTX5000 (48 SMs) and Quadro RTX6000 (72 SMs)~\cite{turing}. DICE is correspondingly scaled by increasing the number of CGRA clusters to match the SM counts (referred to as \textbf{DICE-O48} and \textbf{DICE-O72}). Detailed hardware configurations are summarized in Table~\ref{tab:rtx6000}. For selected kernels, we also increase problem sizes to ensure full SM occupancy.
}

\begin{table}[t]
\centering
\caption{\revorange{Scale-up device configuration: DICE vs. DICE-U configurations (Same total compute resource counts).}}
\label{tab:dice_40pe}
\small
\begin{tabular}{l|c|c}
\toprule
\textbf{Component} & \textbf{DICE} & \textbf{DICE-U} \\
\midrule
CPs/Cluster& 4 CPs/Cluster& 2 CPs/Cluster\\
\midrule
ALUs & 16 PEs/CP, 64/Cluster & 32 PEs/CP, 64/Cluster \\
SFUs & 4/CP, 16/Cluster & 8/CP, 16/Cluster \\
LDST Units & 4/CP, 16/Cluster & 8/CP, 16/Cluster \\
\bottomrule
\end{tabular}
\begin{FlushLeft}
  \footnotesize
  Note: Parameters not shown are identical to those in Table~\ref{tab:system config} in SM.
  \end{FlushLeft}
\end{table}

\begin{table}[t]
\centering
\caption{\revorange{Scale-out device configurations.}}
\label{tab:rtx6000}
\small
\begin{tabular}{C{1.6cm}|C{1.2cm}|C{1.2cm}|C{1.2cm}|C{1.2cm}}
\toprule
\textbf{Component} & \textbf{Quadro RTX5000} & \textbf{DICE-O48} & \textbf{Quadro RTX6000} & \textbf{DICE-O72} \\
\midrule
\# SM or CC & 48& 48& 72& 72\\
\midrule
L2 Cache & \multicolumn{2}{c|}{4096 KB total} & \multicolumn{2}{c}{6144 KB total} \\
DRAM & \multicolumn{2}{c|}{8 channels} & \multicolumn{2}{c}{12 channels} \\
\bottomrule
\end{tabular}
\begin{FlushLeft}
  \footnotesize
  Note: RTX5000/6000 share the same Turing SM as RTX2060S~\cite{turing}; therefore, all SM-level parameters are identical to those in Table~\ref{tab:system config}.
  \end{FlushLeft}
\end{table}

\section{Evaluation Results}
\label{sec:evaluation}

\subsection{RF Accesses}
\begin{figure}[tp!]
    \centering
    \includegraphics[width=\linewidth]{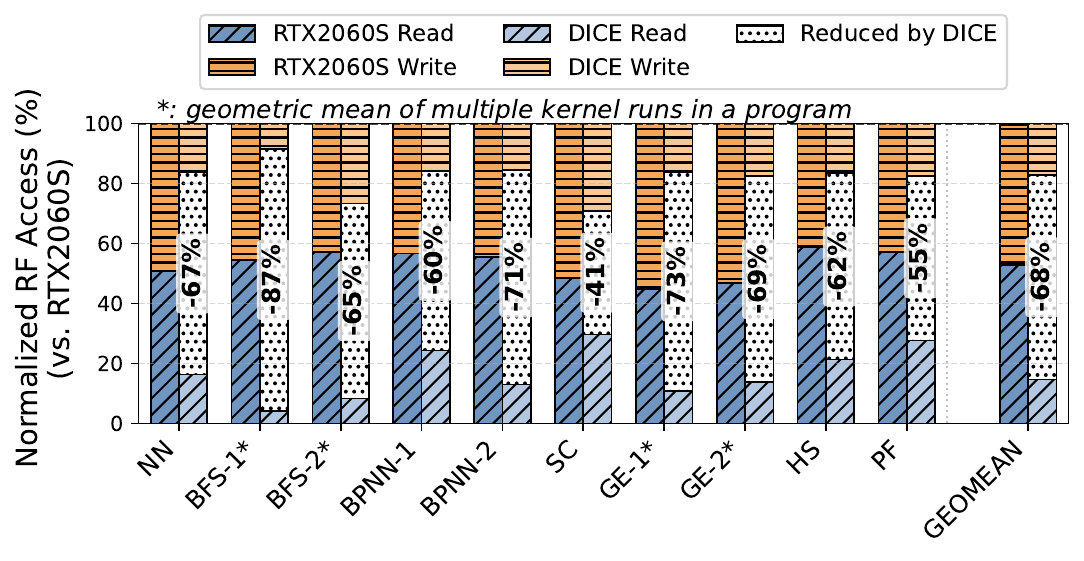}
    \caption{Normalized RF accesses (\%) (DICE vs. RTX2060S)}

    \label{fig:rf ratio}
\end{figure}

Fig.~\ref{fig:rf ratio} shows the normalized RF access counts in DICE relative to RTX2060S and demonstrates that, for the same program, DICE requires only 32\% of the RF accesses compared with RTX2060S on average. This 68\% reduction is primarily attributed to the spatial computing nature of the CGRA, where intermediate data is directly forwarded across wires. Note that thread unrolling and TMCU do not impact RF access counts.

\subsection{Performance}
\begin{figure}[tp!]
    \centering
    \includegraphics[width=\linewidth]{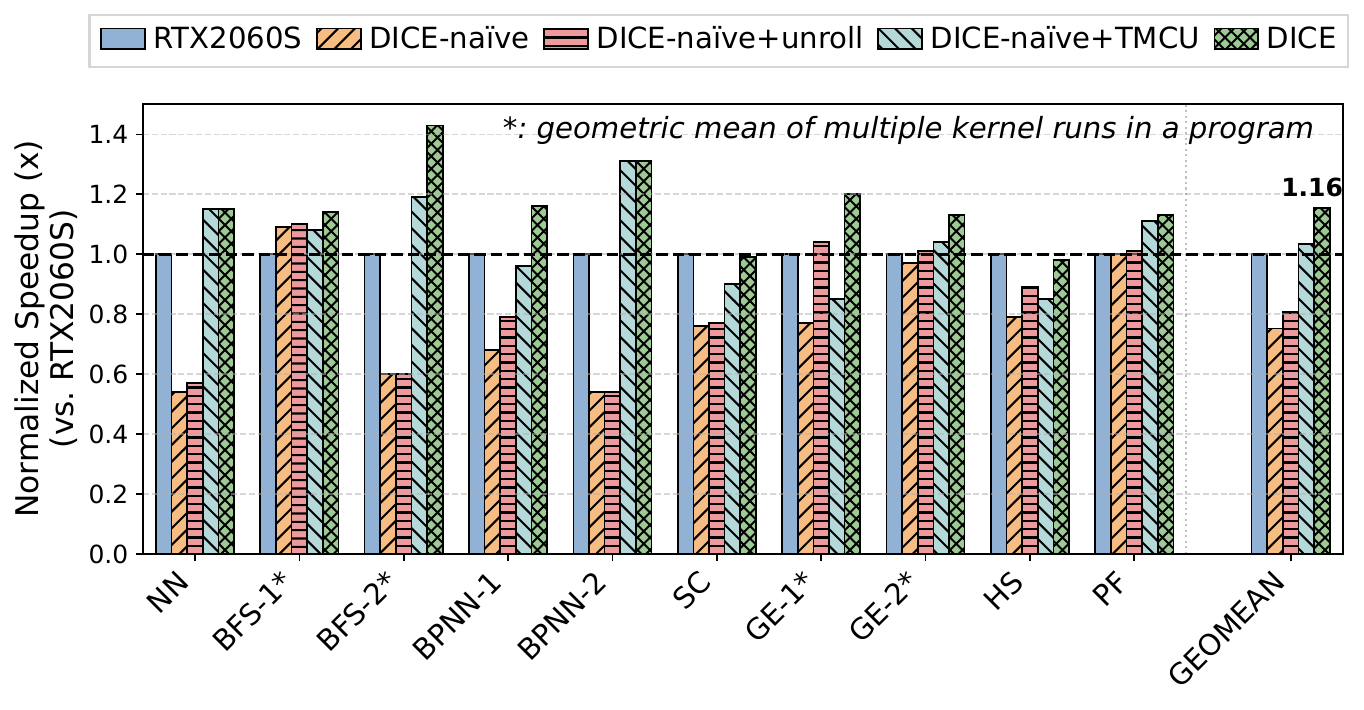}
    \caption{Speedup ($\times$) (DICE variants vs. RTX2060S)}

    \label{fig:speedup}
\end{figure}

\begin{figure}[tp!]
    \centering
    \includegraphics[width=\linewidth]{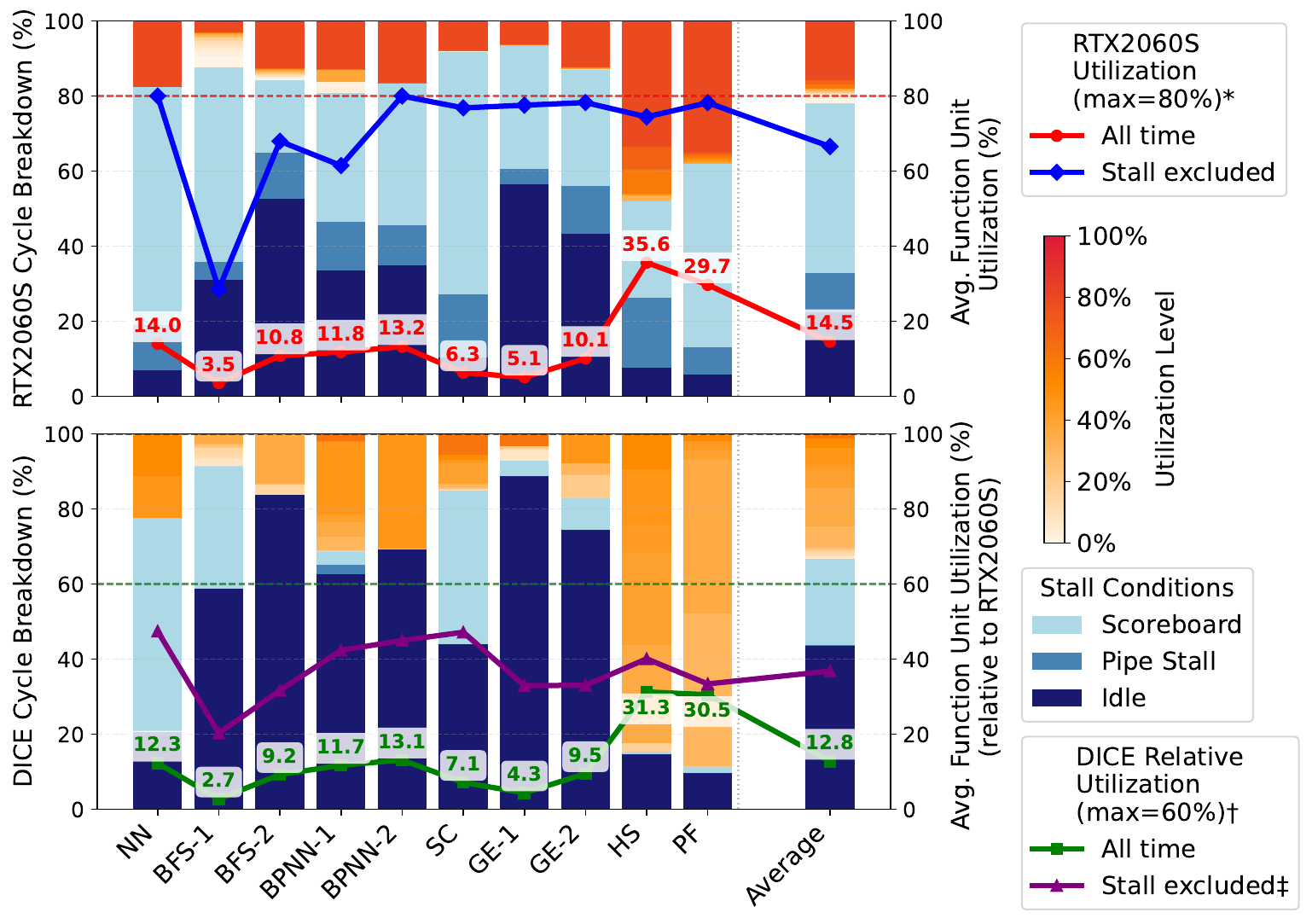}
    \begin{FlushLeft}
    \footnotesize
    \textsuperscript{*}Maximum GPU utilization is 80\% due to dispatcher constraints: each RTX2060S SM contains 160 functional units (64 INT, 64 FP, 16 LD/ST, 16 SFU) but can only dispatch up to 128 threads per cycle.
    
    \textsuperscript{†}DICE relative utilization is capped at 60\%: each CGRA cluster has 96 functional units (4 CPs × 24 functional units) compared to RTX2060S's 160 units per SM.
    
    \textsuperscript{‡}Also represents the average \textit{p-graph} size including memory operations.
    \end{FlushLeft}
    \caption{Cycle breakdown and average functional unit utilization. Top: RTX2060S. Bottom: DICE.}

    \label{fig:pe_utilization}
\end{figure}

Fig.~\ref{fig:speedup} presents the performance comparison between RTX2060S baseline, DICE-na\"ive, DICE-na\"ive+unroll, DICE-na\"ive+TMCU, and DICE as described in Section~\ref{sec:baseline}. Fig.~\ref{fig:pe_utilization} presents a detailed breakdown of cycle distribution and functional unit utilization for both RTX2060S and DICE architectures.

\subsubsection{DICE vs. RTX2060S}
\label{sec:performance_evauluation}
Overall, DICE achieves a geometric mean speedup of 1.16× over RTX2060S, demonstrating that spatial thread-pipeline execution on CGRAs can match conventional SIMD architectures despite fundamental design differences. This performance advantage stems from several architectural trade-offs that we analyze below:

\textbf{Reduced Data Movement Instructions}: The spatial execution model eliminates certain data-movement instructions common in GPGPU workloads—such as \texttt{MOV} (register/value-to-register moves) and \texttt{S2R} (special register moves)—by forwarding values directly between PEs over the CGRA interconnect. This reduces static instruction count by up to 17.8\% in kernels with relatively low per-thread instruction counts (\texttt{NN}, \texttt{BFS-2}, \texttt{GE-1}). Additionally, branch operations are handled in the control pipeline, reducing actual instruction counts per thread. The combined effects explain why DICE can remain performance-competitive with, and in some kernels outperform RTX2060S with lower overall utilization in Fig.~\ref{fig:pe_utilization}. 

\textbf{Selective Dispatch}: Selective dispatch avoids performance loss from GPGPU's mask-based SIMD backend by only dispatching active threads, as evidenced by RTX2060S's lower stall-excluded utilization in divergence-frequent kernels (\texttt{BFS-1}, \texttt{BFS-2}, \texttt{BPNN-1}). 

\textbf{Improved Memory Latency Hiding}: DICE experiences fewer scoreboard stalls (Fig.~\ref{fig:pe_utilization}), reflecting more effective memory latency hiding. This benefit arises from two factors. First, DICE supports 2048 threads per cluster, which doubles the 1024 threads per SM in RTX2060S while maintaining the same register file capacity as RTX2060S (256~KB/SM), thanks to the reduced per-thread register demand enabled by spatial execution. Second, \textit{p-graph} execution naturally spreads memory requests more evenly over time, reducing memory congestion and improving overlap with computation.

\subsubsection{DICE Utilization Analysis} Cycle breakdowns in Fig.~\ref{fig:pe_utilization} show that DICE sustains more active cycles than RTX2060S but at a lower average active utilization, reflecting smaller \textit{p-graph} sizes as indicated by its stall-excluded utilization curve. Utilization varies considerably across kernels. Kernels with regular, compute-heavy codepaths (\texttt{NN}, \texttt{GE-2}) achieve high utilization. Compute-intensive kernels with frequent shared-memory accesses (\texttt{BPNN-1}, \texttt{BPNN-2}, \texttt{HS}, \texttt{PF}) see reduced \textit{p-graph} sizes because shared-memory load latencies cannot be predicted statically due to potential bank conflicts and LDST Unit stalls; frequent synchronization barriers further shorten \textit{p-graphs}. We leave optimizations of shared memory access with CGRA pipeline execution to future work. Kernels such as \texttt{BFS-1}, \texttt{BFS-2}, \texttt{SC}, and \texttt{GE-1} exhibit the lowest utilization due to irregular, data-dependent memory accesses, frequent control divergence, and low computational intensity.

The overall results show that DICE’s architectural advantages yield higher temporal utilization, which offsets its relatively lower spatial utilization and delivers marginally better performance than \shepadd{the modeled} RTX2060S.

\subsubsection{Optimization Effectiveness}
\label{sec:optimization_eval}
\paragraph{Impact of Thread Unrolling}

\texttt{BPNN-1}, \texttt{GE-1}, and \texttt{HS} achieve 1.24--1.35$\times$ speedup over DICE-na\"ive, indicating small \textit{p-graph} sizes without unrolling. Other kernels show little benefit, as their performance is bottlenecked by memory and interconnect bandwidth, or the \textit{p-graphs} are already large enough that unrolling is not needed.

\paragraph{Impact of Temporal Memory Coalescing}

\texttt{NN}, \texttt{BFS-2}, \texttt{BPNN-1}, and \texttt{BPNN-2} achieve 1.41--2.45$\times$ speedup with TMCU only over DICE-na\"ive, showing prior memory/interconnect bandwidth limitations.  
Unlike MSHRs, TMCU merges requests before cache access, producing a single cache access and response that better utilizes cache process bandwidth. Additionally, with
the write-through cache policy used in both DICE and RTX2060S, all write requests are immediately forwarded to the system interconnect, creating congestion. By merging writes before they enter the interconnect, TMCU substantially reduces traffic, alleviates contention, and improves effective memory bandwidth utilization. 

\begin{figure}[t]
    \centering
    \includegraphics[width=\linewidth]{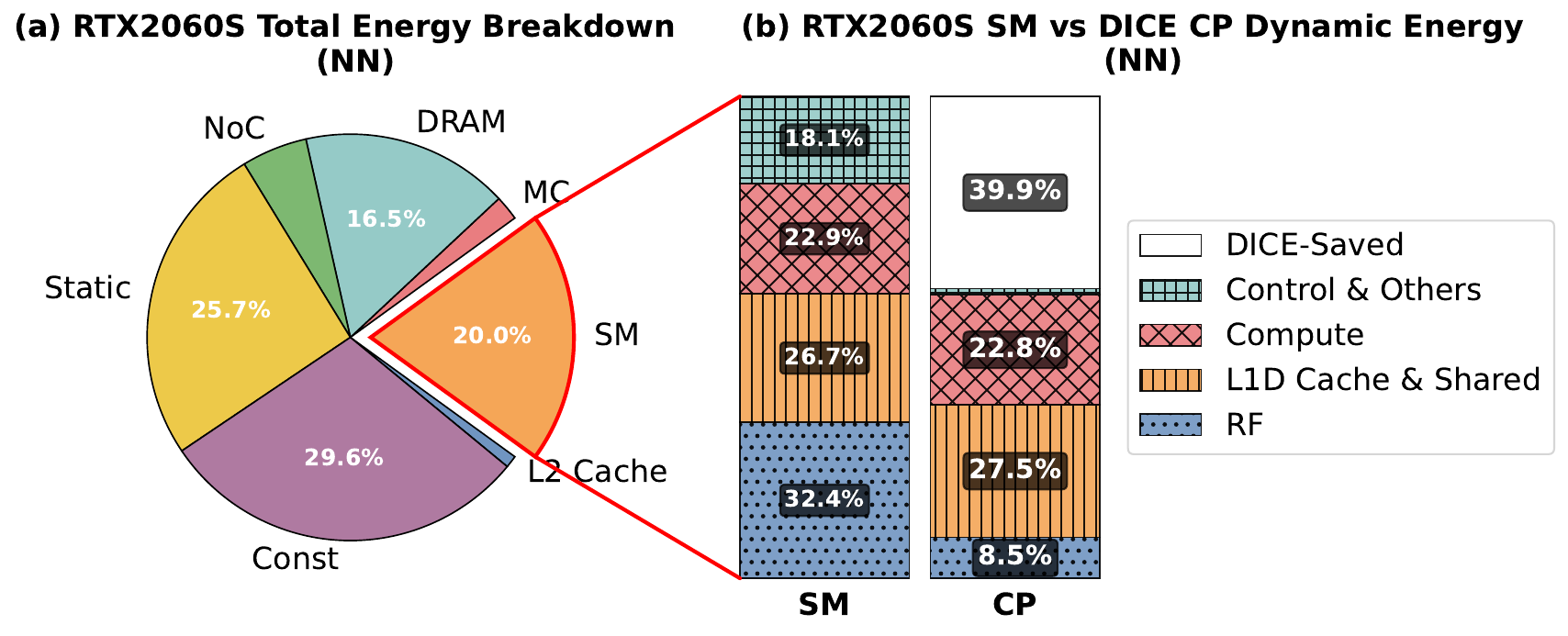}
    \caption{Energy breakdown of NN benchmark in (a) RTX2060S system (b) RTX2060S SMs and DICE CPs (normalized)}

    \label{fig:system_energy}
\end{figure}

\begin{figure}[tp!]
      \centering
      \begin{minipage}[b]{\linewidth}
          \centering
          \includegraphics[width=\linewidth]{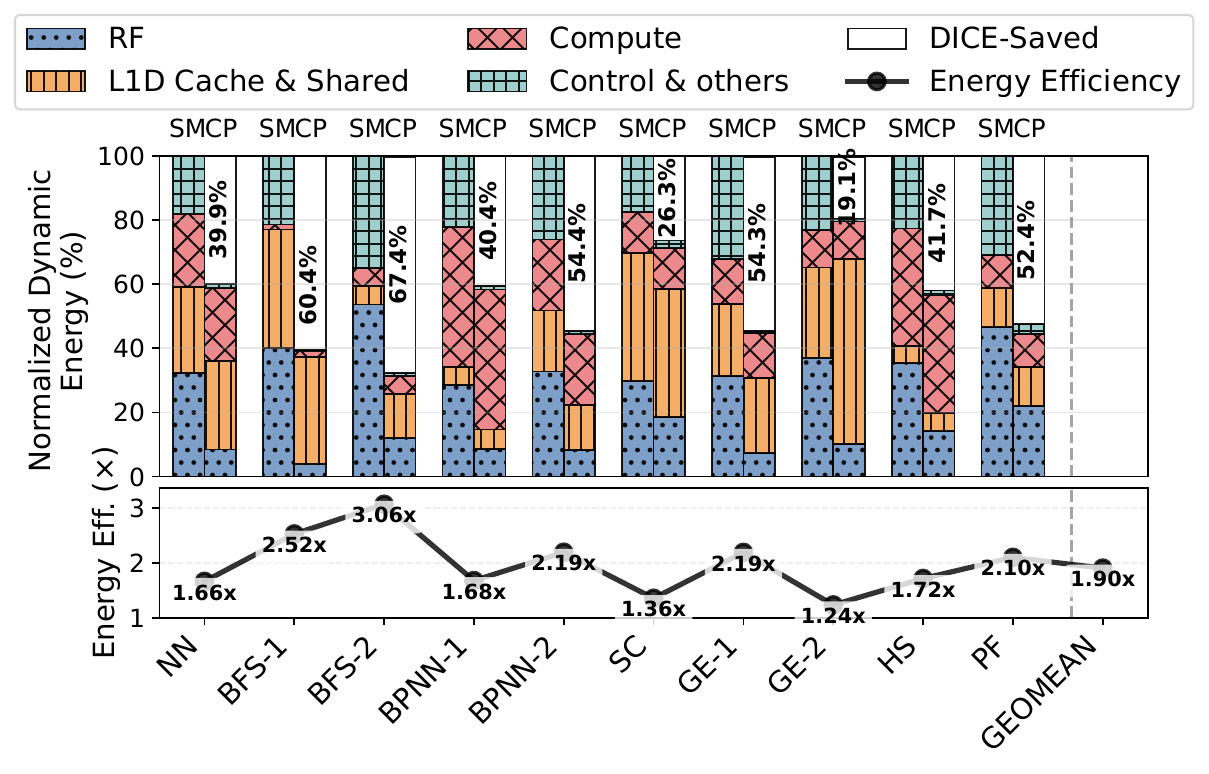}

          {\footnotesize (a) Normalized energy breakdown and energy efficiency ($\times$)}
          \label{fig:breakdown}
      \end{minipage}

      \vspace{1pt}

      \begin{minipage}[b]{\linewidth}
          \centering
          \includegraphics[width=\linewidth]{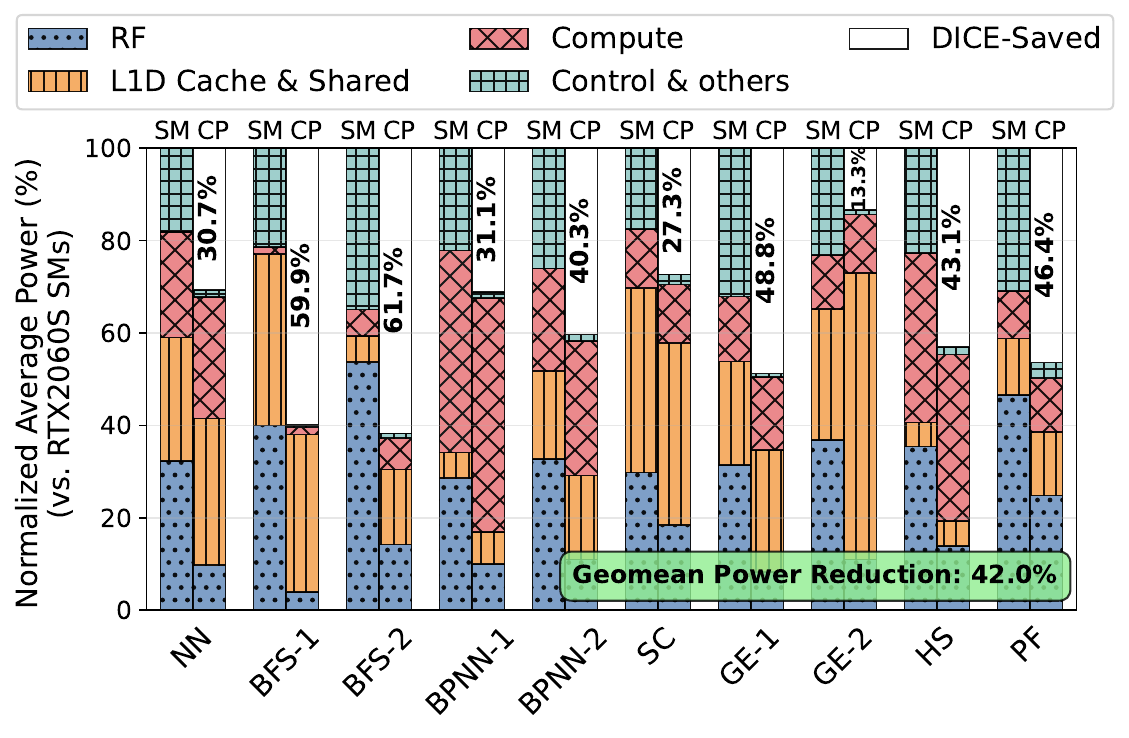}

          {\footnotesize (b) Normalized average dynamic power consumption}
          \label{fig:power_efficiency}
      \end{minipage}

      \caption{Energy \& power efficiency (DICE CPs vs. RTX2060S SMs)}
      \label{fig:energy_power_comparison}
  \end{figure}
  
\paragraph{Combined Optimizations and Performance Saturation}
For kernels such as \texttt{BFS-2}, \texttt{BPNN-1}, \texttt{SC}, \texttt{GE-1}, \texttt{GE-2}, and \texttt{HS}, DICE significantly outperforms either unrolling or TMCU alone. Unrolling raises resource utilization but also increases memory request rates, shifting the bottleneck to memory and interconnect bandwidth. This synergy yields a 1.54$\times$ speedup over the DICE-na\"ive. In contrast, other kernels show negligible to modest improvement from unrolling after TMCU, as memory bandwidth is already saturated. These results indicate that DICE can both alleviate architectural inefficiencies and, in some cases, drive performance to the hardware’s fundamental limits.

\subsection{Energy Efficiency and Power}

We evaluate energy efficiency and power consumption between DICE and RTX2060S. Fig.~\ref{fig:system_energy}(a) presents the system-level energy distribution of RTX2060S on \texttt{NN} benchmark, breaking down consumption across Streaming Multiprocessors (SMs), Network-on-Chip (NoC), L2 caches, memory controllers (MCs), idle cores, off-chip DRAM, and auxiliary circuits (CONST). The left bar chart in Fig.~\ref{fig:system_energy}(b) further decomposes the SM dynamic energy into individual parts. Since DICE primarily replaces GPGPU SMs with CPs, and separating static energy in AccelWattch~\cite{accelwattch} is challenging, we focus on the dynamic energy consumed in SMs and CPs. 

The right bar chart in Fig.~\ref{fig:system_energy}(b) illustrates the normalized energy consumption breakdown of the DICE CP components relative to the RTX2060S SMs on \texttt{NN} benchmark. The white portion (39.9\%) represents the total energy savings achieved by the CPs. These savings come from multiple factors. First, reduced RF accesses substantially lower overall energy use from 32.4\% to 8.5\%; Second, DICE's lightweight control pipeline reduces control logic energy from 18.1\% to 1.3\% by eliminating warp-level control logic and amortizing remaining control overhead across hundreds of threads.

L1 data cache and shared memory accesses also consume over 26.7\% of SM dynamic energy. In DICE CPs, this energy is comparable to the RTX2060S SMs thanks to the effective coalescing provided by the TMCU, underscoring the importance of memory coalescing. Without it, the increased cache accesses would substantially raise energy consumption. However, even with the TMCU, CP's L1 energy is slightly higher than SMs in some cases, as the TMCU's effectiveness diminishes when coalescable requests arrive non-contiguously or when thread dispatching stalls.

Fig.~\ref{fig:energy_power_comparison}(a) extends this analysis across all benchmarks.
DICE CPs achieve a geometric mean of 1.90× energy efficiency compared to RTX2060S SMs. An exception occurs in \texttt{GE-2}, where DICE spends more energy on L1D cache and shared memory due to a load from the same address across all threads. This overhead could be eliminated by placing the data in the shared constant buffer reusable across all threads.

Fig.~\ref{fig:energy_power_comparison}(b) further shows that DICE reduces average power consumption by 42.0\%. This average power reduction closely tracks the energy savings because DICE's performance gains are modest (1.16× on average), meaning the energy efficiency improvements translate almost directly to power savings. This substantial power reduction is particularly valuable for thermally constrained environments and enables higher computational density within fixed power budgets.

\subsection{Area Comparison}
\label{sec:area}

\begin{figure}[tp!]
    \centering
    \includegraphics[width=\linewidth]{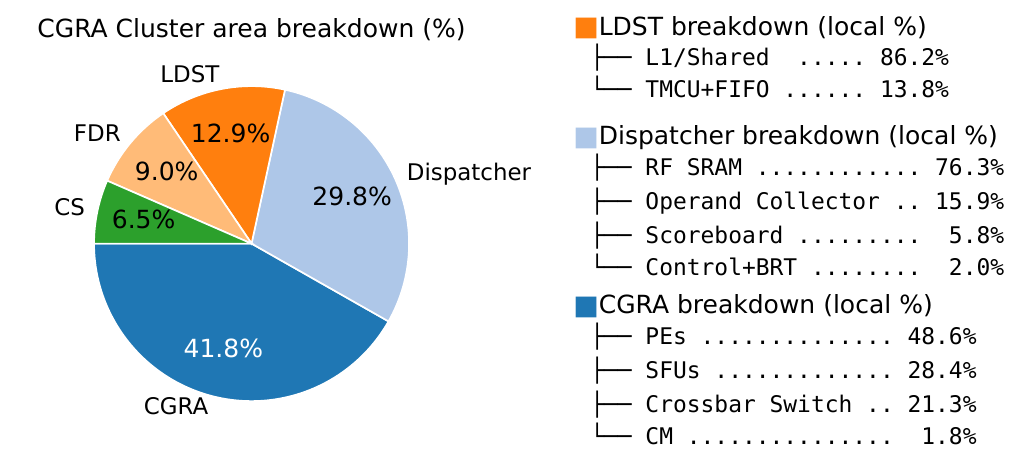}
    \caption{DICE area breakdown}
    \label{fig:area_breakdown}
\end{figure}

\shepadd{
To estimate area, we implement the DICE-specific RTL components and synthesize them in FreePDK45~\cite{freepdk45}, while modeling SRAM/cache structures with CACTI~\cite{cacti}. The total area of a DICE CGRA cluster is 16.21 mm$^2$, with the detailed breakdown shown in Fig.~\ref{fig:area_breakdown}. Because public area breakdowns for commercial GPUs are limited, we first provide a qualitative comparison against an RTX2060S SM, then give a conservative quantitative estimate based on die-shots.
}

\shepadd{
\textbf{Qualitative analysis}: We divide DICE's area into four categories. $A_1$: GPU-matched components, including compute units and SRAM structures, whose total area should be broadly comparable to that of an RTX 2060 Super SM. DICE's unified INT/FP datapath may further reduce PE area by up to
30\%~\cite{fused_alu}, though we do not include this benefit in the estimate below. $A_2$: DICE-specific components, including the CGRA switches and configuration memory, account for 9.7\% of a DICE CP's area. $A_3$: modified components, including the PDOM-stack, operand collector, scoreboard, control logic, and TMCU, account for 12.1\% of a DICE CP's area. $A_4$: removed GPU components include the instruction buffer, warp scheduler, and warp-related control logic. The total core area can therefore be written as
$A_{\text{DICE}} = A_1 + A_2 + A_{3,\text{DICE}} - A_{3,\text{GPU}} - A_4$.
We assume $A_{3,\text{DICE}} \approx A_{3,\text{GPU}}$, as DICE modifies but does not fundamentally scale these structures, and conservatively set $A_4 = 0$. This yields $A_{\text{GPU}} \approx A_1 + A_{3,\text{GPU}}$, and thus the relative upper-bound area overhead can be expressed as $A_{\text{DICE}}/{A_{\text{GPU}}} - 1 = 10.7\%$.}

\shepadd{
\textbf{Estimated quantitative comparison}: We use die shots of the RTX 2060 Super~\cite{fritz_tu106_2019} and GTX1660Ti~\cite{fritz_tu116_2019,nvidia_gtx1660ti_tu116}, both 12nm Turing GPUs, where the latter's SM excludes Tensor Cores and RT cores. From these, we estimate an SM area of 5.44 mm$^2$ for the RTX2060S and 4.46 mm$^2$ for the GTX 1660Ti. In comparison, a DICE CGRA cluster occupies 2.92 mm$^2$ after conservative scaling to 12nm~\cite{scale}. Although this is still an approximate comparison, it suggests that a DICE CGRA cluster is in the same general area range as, if not smaller than, a full GTX 1660Ti SM.
}




\subsection{Scalability Evaluation}
\label{sec:scale_result}

\begin{figure}[tp!]
      \centering
      \begin{minipage}[b]{0.48\linewidth}
          \centering
          \includegraphics[width=\linewidth]{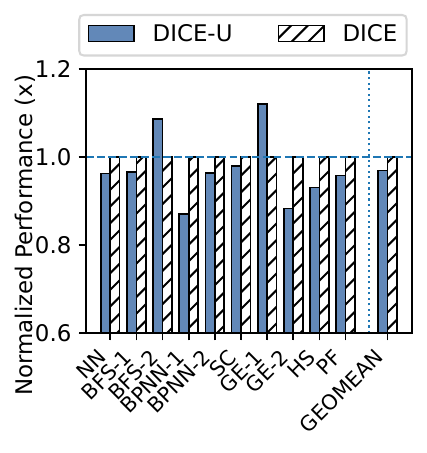}

          {\footnotesize (a) Normalized performance ($\times$)}
      \end{minipage}
      \hfill
      \begin{minipage}[b]{0.48\linewidth}
          \centering
          \includegraphics[width=\linewidth]{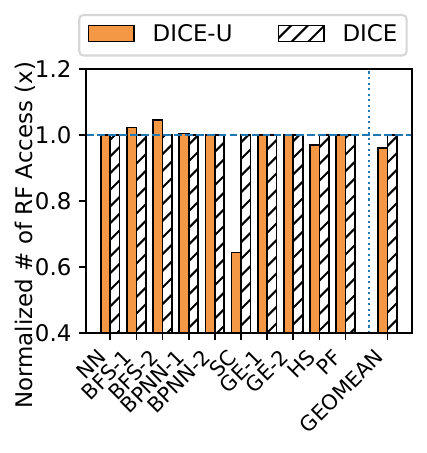}

          {\footnotesize (b) Normalized RF Access ($\times$)}
      \end{minipage}

      \caption{\revorange{Normalized performance and number of RF accesses (DICE-U vs. DICE)}}
      \label{fig:scale-up}
  \end{figure}

\begin{figure}[tp!]
    \centering
    \includegraphics[width=\linewidth]{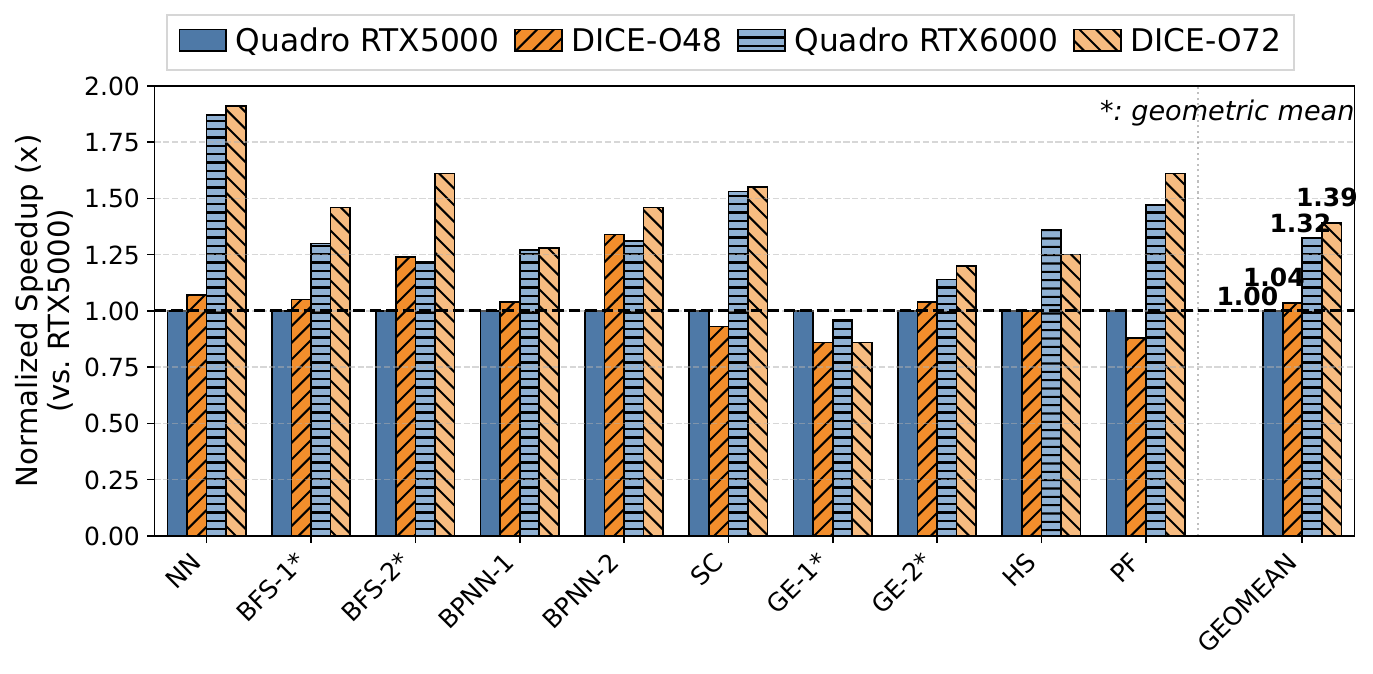}
    \caption{\revorange{Speedup ($\times$) of DICE-O48, DICE-O72, and RTX6000 (vs. RTX5000).}}
    \label{fig:scale-out-speedup}
\end{figure}

\begin{figure}[tp!]
      \centering

      \begin{minipage}[b]{0.49\linewidth}
          \centering
          \includegraphics[width=\linewidth]{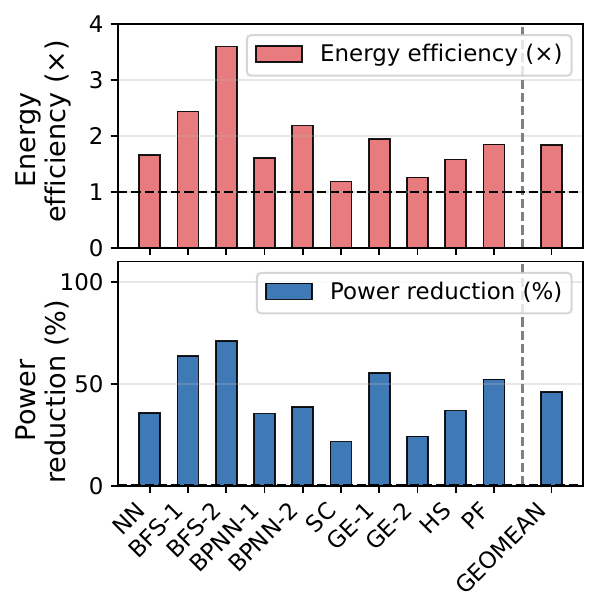}
          \par\vspace{2pt}
          {\footnotesize (a) DICE-O48 vs. RTX5000}
      \end{minipage}
      \hfill
      \begin{minipage}[b]{0.49\linewidth}
          \centering
          \includegraphics[width=\linewidth]{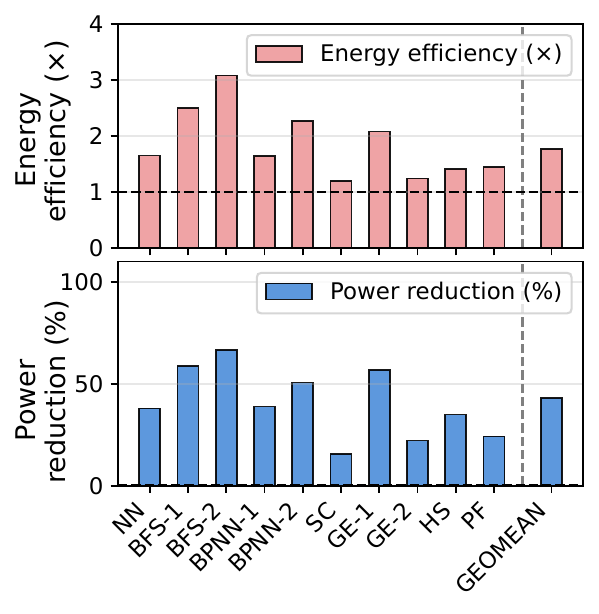}
          \par\vspace{2pt}
          {\footnotesize (b) DICE-O72 vs. RTX6000}
      \end{minipage}

      \caption{\revorange{Dynamic energy efficiency and power reduction (DICE CPs vs. Turing SMs)}}
      \label{fig:scale-out-energy}
  \end{figure}

\subsubsection{Scale-up (Larger CGRAs)}
As shown in Fig.~\ref{fig:scale-up}, substituting larger CGRAs while keeping the total resource the same degrades the performance slightly (0.97$\times$) due to more underutilized PEs, as some workloads cannot fully exploit the enlarged spatial fabric. On the positive side, RF accesses decrease by approximately 3.8\% overall, since larger CGRAs enable mapping of bigger \textit{p-graphs}, particularly in compute-intensive kernels such as \texttt{SC} and \texttt{HS}. However, for control-divergent workloads such as \texttt{BFS} and \texttt{GE-1}, RF accesses slightly increase. This is because merged \textit{p-graphs} introduce more predicated execution, causing registers of masked-off threads to be read even though their results are eventually discarded. Overall, scale-up is inherently limited. The exploitable \textit{p-graph} size of most workloads is constrained by memory dependences and control divergence, and larger CGRAs increase compiler and mapping complexity. Thus, simply enlarging the CGRA is not a scalable path for performance improvement.

\subsubsection{Scale-out (More CPs)}
Fig.~\ref{fig:scale-out-speedup} and Fig.~\ref{fig:scale-out-energy} compare DICE against larger Turing GPUs (Quadro RTX5000 and RTX6000).
The corresponding DICE variants proportionally scale the number of CPs to match the compute capacity of these GPUs. The results are consistent with our earlier evaluation on RTX2060S: DICE achieves performance \shepreplace{marginally higher than}{comparable to the} \shepreplace{scaled Turing GPUs}{scaled Turing GPU baselines} (1.04$\times$--1.05$\times$), while delivering substantially better energy efficiency (1.77$\times$--1.83$\times$) and significant average power reductions (43.0\%--45.9\%). Overall, these results demonstrate that DICE preserves the desirable scale-out property of GPUs. Performance scales effectively by increasing the number of CPs, while maintaining superior energy efficiency and power advantages.

\subsubsection{Performance Comparison Against Newer GPUs}
\begin{figure}[tp!]
      \centering

      \begin{minipage}[b]{0.49\linewidth}
          \centering
          \includegraphics[width=\linewidth]{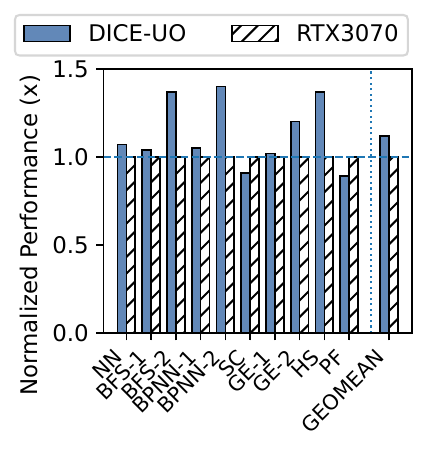}
          \par\vspace{2pt}
          {\footnotesize (a) Speedup ($\times$)}
      \end{minipage}
      \hfill
      \begin{minipage}[b]{0.49\linewidth}
          \centering
          \includegraphics[width=\linewidth]{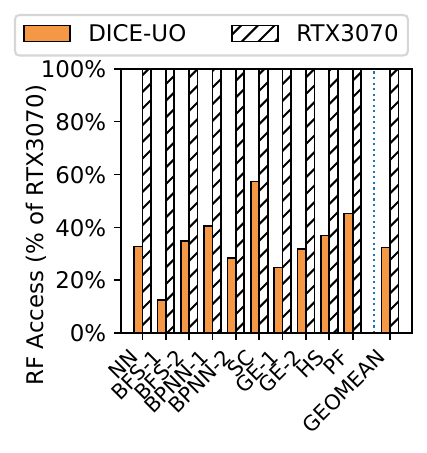}
          \par\vspace{2pt}
          {\footnotesize (b) RF Access (\%)}
      \end{minipage}

      \caption{\revorange{Speedup ($\times$) and RF Accesses Ratio (\%) (DICE-UO vs. RTX3070)}}
      \label{fig:3070_result}
  \end{figure}

We additionally compare DICE against the RTX3070.
Since each RTX3070 SM offers 2$\times$ the FP32 throughput of a Turing SM, we scale DICE both up and out (\textbf{DICE-UO}) to match the RTX3070 compute resource count, as summarized in Table~\ref{tab:dice_3070}. As shown in Fig.~\ref{fig:3070_result}, DICE maintains comparable average performance while reducing RF accesses by 68\%, indicating that DICE’s efficiency carries over to newer GPU baselines.

\begin{table}[t]
\centering
\caption{\revorange{DICE-UO vs. RTX3070 Configuration (1132 MHz)}}
\label{tab:dice_3070}
\small
\begin{tabular}{C{2.5cm}|C{2.5cm}|C{2.5cm}}
\toprule
\textbf{Component} & \textbf{RTX3070~\cite{3070}} & \textbf{DICE-UO}  \\
\midrule
\# SM or CC & 46 & 46\\
Core Partition& 4 subcore/SM  & 4 CPs/Cluster\\
\midrule
CUDA Cores & 32\textsuperscript{*}/subcore & 32 PEs/CP \\
\midrule
L1 Cache/Shared & \multicolumn{2}{c}{128 KB/(SM or CC)} \\
\bottomrule
\end{tabular}
\begin{FlushLeft}
  \footnotesize
  \textbf{Note:} Parameters not listed here are identical to those in Table~\ref{tab:system config} at SM/Cluster level. 
  \textsuperscript{*}Half of the CUDA cores support FP32/INT32, while the other half support FP32 only.
\end{FlushLeft}
\end{table}

\section{Related Work}
\label{sec:related work}
\textbf{CGRA Memory Handling}:
Prior CGRAs handle variable memory latency through three main approaches. Scratchpad-based designs (HyCube~\cite{hycube}, OpenCGRA~\cite{opencgra}, AHA~\cite{AHA}) restrict memory operations to fixed-latency buffers. However, in massively parallel SIMT programs found in real-world applications, when the memory access pattern is unknown at compile time, these buffers must be sized for worst-case demand, potentially requiring DRAM-scale storage that is impractical on-chip. Flow-controlled CGRAs (Plasticine~\cite{plasticine}, SNAFU~\cite{snafu}, Riptide~\cite{riptide}) use handshaking protocols between PEs to tolerate variable latency. However, PEs stall when waiting for operands, potentially blocking for hundreds of cycles on cache misses. Dynamic dataflow architectures (SGMF~\cite{sgmf}, VGIW~\cite{vgiw}, dMT-CGRA~\cite{dmt-cgra}) employ a tagged-token mechanism to achieve non-blocking, out-of-order processing, but at the cost of substantial hardware overhead. 
In contrast, DICE's \textit{p-graph} partitioning sidesteps these issues by separating memory consumers from producers to different configurations at compile time, preserving a cache-based memory system without the need for elastic interconnect between PEs.

\textbf{CGRA for SIMT}:
CGRA has been previously applied to the SIMT model. SGMF~\cite{sgmf} uses a dynamically scheduled dataflow CGRA with tagged tokens to pipeline multiple threads through a shared dataflow graph. VGIW~\cite{vgiw} extends SGMF by introducing control flow coalescing and also a von Neumann-style control to orchestrate basic block execution.
dMT-CGRA~\cite{dmt-cgra} further adds fine-grained inter-thread communication directly through the CGRA fabric.
As described before, they all use dynamic tagged-token CGRAs instead of minimal-overhead statically scheduled CGRAs. DICE applies \textit{p-graph} partitioning and further introduces thread unrolling and the TMCU, which are not explored in these prior works.

\textbf{Decoupled CGRA Execution and Memory Operation}:
Softbrain~\cite{stream_dataflow} separates memory operations from computation in streaming dataflow architectures but focuses on stream access patterns rather than general-purpose SIMT execution. DICE adopts a different approach with \textit{p-graphs} where the CGRA generates memory requests and forwards them directly to the load/store unit, supporting arbitrary access patterns.

\textbf{More CGRAs/Processor Integration}:
ADRES~\cite{adres} integrates a CGRA with a VLIW processor, sharing functional units and register files between the two modes. 
DySER~\cite{dyser} tightly couples a CGRA into a CPU pipeline.
TRIPS~\cite{trips} employs a tiled architecture that composes simple cores into larger logical units, targeting both instruction-level and thread-level parallelism. 
WaveScalar~\cite{wavescalar} is a dataflow architecture that distributes execution across a grid of processing elements. 
All these systems explore different ways to integrate CGRAs into CPU-like environments, focusing on composability, dynamic specialization, or single-thread performance.
In contrast, DICE is tailored to harness massive thread-level parallelism and memory patterns typical in GPGPU workloads.

\section{Summary}
\label{sec:summary}
We presented DICE, an architecture that adopts statically scheduled CGRAs to address critical energy inefficiencies in conventional GPGPUs. 
Our evaluation shows DICE CPs achieve a geometric mean of 1.77--1.90$\times$ dynamic energy efficiency, 42.0\%--45.9\% dynamic power reduction relative to NVIDIA Turing SMs, and performance \shepreplace{marginally better than}{comparable to the modeled} NVIDIA Turing \shepreplace{GPGPUs}{GPU baselines} 
\shepreplace{with equivalent resources}{under matched compute-resource counts and total cache/SRAM capacity}, providing a practical path forward for efficient SIMT computing.

\section*{Acknowledgment}
We thank the anonymous reviewers and the shepherd for their insightful comments and suggestions. We also thank the DICE RTL team members—Aadithya Manoj, Albert Ton, Elliot Norman, Juwon Jun, Patrick Howe, Rambod Taherian, and Sibo Zhang—for their contributions to the development of the DICE RTL. We also used ChatGPT for light editorial assistance to improve sentence clarity and readability.

\appendix
\section{Artifact Appendix}

\subsection{Abstract}
This artifact contains the code, benchmark inputs, DICE metadata bundles, and plotting scripts used to evaluate DICE against baseline GPGPU-Sim. It supports three experiment groups from the paper: base comparison, scale-up, and scale-out. The intended artifact-evaluation path is a full rerun of the experiments followed by plot generation.

\subsection{Artifact Checklist (Meta-Information)}

{\small
\begin{itemize}
  \item \textbf{Program:} Bash scripts, Python scripts, CUDA Rodinia benchmarks, GPGPU-Sim
  \item \textbf{Compilation:} \texttt{make}, \texttt{gcc}/\texttt{g++}, \texttt{nvcc}
  \item \textbf{Data set:} Bundled Rodinia inputs
  \item \textbf{Run-time environment:} 64-bit Linux
  \item \textbf{Hardware:} Multicore CPU; no physical GPU required
  \item \textbf{Metrics:} Speedup; normalized RF accesses
  \item \textbf{Output:} CSV, PDF, PNG
  \item \textbf{Experiments:} Base; scale-up; scale-out
  \item \textbf{How much disk space required (approximately)?} About 15--20~GB
  \item \textbf{How much time is needed to prepare workflow (approximately)?} About 15--30 minutes
  \item \textbf{How much time is needed to complete experiments (approximately)?} Several hours for a full rerun
  \item \textbf{Publicly available?} Yes
  \item \textbf{Workflow automation framework used:} Bash, \texttt{make}, and Python
  \item \textbf{Archived (provide DOI):} \url{https://doi.org/10.5281/zenodo.19278715}
\end{itemize}
}

\subsection{Description}

\subsubsection{How to Access}
The archived artifact is available at \url{https://doi.org/10.5281/zenodo.19278715}. A source mirror is available at \url{https://github.com/jiayi-wang98/DICE-test-collection}. If the GitHub repository is used instead of the archive, initialize the submodules before running anything:

 \begin{verbatim}
  git clone https://github.com/\
  jiayi-wang98/\
  DICE-test-collection.git
  cd DICE-test-collection
  git submodule update --init --recursive
  \end{verbatim}

\subsubsection{Hardware Dependencies}
A standard 64-bit Linux machine is sufficient. A physical NVIDIA GPU is not required because the experiments are simulation-based. A multicore CPU is recommended because the scripts use parallel builds by default.

\subsubsection{Software Dependencies}
The artifact requires \texttt{bash}, GNU \texttt{make}, \texttt{gcc}/\texttt{g++}, and the NVIDIA CUDA Toolkit 11.7 (including \texttt{nvcc} and \texttt{ptxas}). It also requires Python~3 with \texttt{numpy} and \texttt{matplotlib}. The practical Linux package dependencies are the usual GPGPU-Sim requirements: \texttt{build-essential}, \texttt{xutils-dev}, \texttt{bison}, \texttt{flex}, \texttt{zlib1g-dev}, \texttt{libglu1-mesa-dev}, \texttt{libxi-dev}, \texttt{libxmu-dev}, and \texttt{libglut3-dev}. In practice, a compatible \texttt{libasan} runtime is also required because several benchmarks are built with AddressSanitizer. No proprietary EDA tools are required.

\subsubsection{Data Sets}
All benchmark inputs needed by the automated workflow are already bundled in the artifact under \path{dice-test-gpu-rodinia/data} and \path{gpu-rodinia/data}. The DICE runs also require the bundled metadata/PPTX directories \path{dice-test-gpu-rodinia/cuda/dice_test/sw_*}.

\subsubsection{Models}
The artifact does not contain a machine-learning model. The relevant configuration models are the included simulator configuration files for the evaluated RTX2060S-, RTX3070-, RTX5000-, and RTX6000-style systems.

\subsection{Installation}
Run all commands from the repository root. Set the CUDA paths before running the workflow:

\begin{verbatim}
export CUDA_INSTALL_PATH=\
/usr/local/cuda-11.7
export PTXAS_CUDA_INSTALL_PATH=\
/usr/local/cuda-11.7
\end{verbatim}

On Ubuntu-like systems, the required host packages can be installed with:

\begin{verbatim}
sudo apt-get install git xutils-dev bison \
  flex zlib1g-dev libglu1-mesa-dev \
  libxi-dev libxmu-dev libasan5 python3 \
  python3-numpy python3-matplotlib
\end{verbatim}

On RHEL/Rocky/Alma-like systems, the equivalent packages can be installed with:

\begin{verbatim}
sudo dnf install git gcc gcc-c++ make \
  bison flex zlib-devel \
  mesa-libGLU-devel libXi-devel \
  libXmu-devel freeglut-devel \
  python3 python3-numpy python3-matplotlib 
\end{verbatim}

If the AddressSanitizer runtime is missing on those systems, install \texttt{libasan}.

If the local CUDA installation path differs from \path{/usr/local/cuda-11.7}, update both \path{dice-test-gpu-rodinia/common/make.config} and \path{gpu-rodinia/common/make.config} accordingly.

\subsection{Experiment Workflow}
The intended artifact-evaluation path is a full rerun:

A prebuilt Docker image is also available at \texttt{jiayiwang0710/dice-isca-eval:cuda11.7}. Reviewers can pull the image and enter an interactive shell with:

\begin{Verbatim}[breaklines=true, breakanywhere=true]
docker pull jiayiwang0710/dice-isca-eval:cuda11.7
docker run --rm -it jiayiwang0710/dice-isca-eval:cuda11.7 shell
\end{Verbatim}

Inside the container, run:

\begin{verbatim}
bash integrated_test/run_base_test.sh
bash integrated_test/run_scale_out_test.sh
\end{verbatim}

The first script builds and runs the base and scale-up experiments. The second script builds and runs the scale-out experiments. Generated outputs are written to \path{integrated_test/generated_base}, \path{integrated_test/generated_scale_up}, and \path{integrated_test/generated_scale_out}.

\subsection{Evaluation and Expected Results}
The main generated PDFs corresponding to the paper figures are:

\begin{itemize}
  \item Figure~9: \path{integrated_test/generated_base/rf_access_reduction.pdf}
  \item Figure~10: \path{integrated_test/generated_base/dice_gpu_speedup.pdf}
  \item Figure~15: \path{integrated_test/generated_scale_up/40PE_vs_20PE_perf.pdf} and \path{integrated_test/generated_scale_up/40PE_vs_20PE_rf.pdf}
  \item Figure~16: \path{integrated_test/generated_scale_out/scale_out_speedup.pdf}
  \item Figure~18: \path{integrated_test/generated_scale_out/scale_out_3070_speedup.pdf} and \path{integrated_test/generated_scale_out/scale_out_3070_rf_reduction.pdf}
\end{itemize}

The corresponding summary CSV files are stored in the same output directories.

\subsection{Experiment Customization}
Reviewers can override \texttt{JOBS=<n>} to control parallelism, and can selectively enable or disable parts of the workflow using the Boolean environment variables already exposed by the two top-level wrapper scripts.

\subsection{Notes}
This appendix documents the fresh-machine full-rerun path. It intentionally does not rely on regenerating plots from pre-existing results.


\bibliographystyle{IEEEtranS}
\bibliography{refs}

@String{Computing = "Computing" }

@String{Computer = "{IEEE} Computer" }

@String{Springer = "Springer-Verlag" }

@inproceedings{Hong_Kim_2010,
author = {Hong, Sunpyo and Kim, Hyesoon},
title = {An integrated {GPU} power and performance model},
year = {2010},
isbn = {9781450300537},
publisher = {Association for Computing Machinery},
address = {New York, NY, USA},
url = {https://doi.org/10.1145/1815961.1815998},
doi = {10.1145/1815961.1815998},
booktitle = {Proceedings of the 37th Annual International Symposium on Computer Architecture},
pages = {280–289},
numpages = {10},
location = {Saint-Malo, France},
series = {ISCA '10}
}

@inproceedings{gpuwattch,
author = {Leng, Jingwen and Hetherington, Tayler and ElTantawy, Ahmed and Gilani, Syed and Kim, Nam Sung and Aamodt, Tor M. and Reddi, Vijay Janapa},
title = {{GPUWattch}: enabling energy optimizations in {GPGPUs}},
year = {2013},
isbn = {9781450320795},
publisher = {Association for Computing Machinery},
address = {New York, NY, USA},
url = {https://doi.org/10.1145/2485922.2485964},
doi = {10.1145/2485922.2485964},
booktitle = {Proceedings of the 40th Annual International Symposium on Computer Architecture},
pages = {487–498},
numpages = {12},
location = {Tel-Aviv, Israel},
series = {ISCA '13}
}

@book{gpgpu_book, series={Synthesis Lectures on Computer Architecture}, title={General-Purpose Graphics Processor Architectures}, ISBN={978-3-031-00631-9}, DOI={10.1007/978-3-031-01759-9}, address={Cham}, publisher={Springer}, author={Aamodt, Tor M. and Fung, Wilson Wai Lun and Rogers, Timothy G.}, year={2018}, edition={1}, pages={122} }

@ARTICLE{nv_tesla,
  author={Lindholm, Erik and Nickolls, John and Oberman, Stuart and Montrym, John},
  journal={IEEE Micro}, 
  title={{NVIDIA Tesla}: A Unified Graphics and Computing Architecture},
  year={2008},
  volume={28},
  number={2},
  pages={39-55},
  doi={10.1109/MM.2008.31}}

@INPROCEEDINGS{Fung_Sham_Yuan_Aamodt_2007,
  author={Fung, Wilson W.L. and Sham, Ivan and Yuan, George and Aamodt, Tor M.},
  booktitle={40th Annual IEEE/ACM International Symposium on Microarchitecture (MICRO 2007)}, 
  title={Dynamic Warp Formation and Scheduling for Efficient {GPU} Control Flow},
  year={2007},
  volume={},
  number={},
  pages={407-420},
  doi={10.1109/MICRO.2007.30}}

@article{AHA,
author = {Koul, Kalhan and Melchert, Jackson and Sreedhar, Kavya and Truong, Leonard and Nyengele, Gedeon and Zhang, Keyi and Liu, Qiaoyi and Setter, Jeff and Chen, Po-Han and Mei, Yuchen and Strange, Maxwell and Daly, Ross and Donovick, Caleb and Carsello, Alex and Kong, Taeyoung and Feng, Kathleen and Huff, Dillon and Nayak, Ankita and Setaluri, Rajsekhar and Thomas, James and Bhagdikar, Nikhil and Durst, David and Myers, Zachary and Tsiskaridze, Nestan and Richardson, Stephen and Bahr, Rick and Fatahalian, Kayvon and Hanrahan, Pat and Barrett, Clark and Horowitz, Mark and Torng, Christopher and Kjolstad, Fredrik and Raina, Priyanka},
title = {{AHA}: An Agile Approach to the Design of Coarse-Grained Reconfigurable Accelerators and Compilers},
year = {2023},
issue_date = {March 2023},
publisher = {Association for Computing Machinery},
address = {New York, NY, USA},
volume = {22},
number = {2},
issn = {1539-9087},
url = {https://doi.org/10.1145/3534933},
doi = {10.1145/3534933},
journal = {ACM Trans. Embed. Comput. Syst.},
month = jan,
articleno = {35},
numpages = {34}
}

@INPROCEEDINGS{accelsim,
  author={Khairy, Mahmoud and Shen, Zhesheng and Aamodt, Tor M. and Rogers, Timothy G.},
  booktitle={2020 ACM/IEEE 47th Annual International Symposium on Computer Architecture (ISCA)}, 
  title={{Accel-Sim}: An Extensible Simulation Framework for Validated {GPU} Modeling},
  year={2020},
  volume={},
  number={},
  pages={473-486},
  doi={10.1109/ISCA45697.2020.00047}}

@article{dyser,
  author={Govindaraju, Venkatraman and Ho, Chen-Han and Nowatzki, Tony and Chhugani, Jatin and Satish, Nadathur and Sankaralingam, Karthikeyan and Kim, Changkyu},
  journal={IEEE Micro}, 
  title={{DySER}: Unifying Functionality and Parallelism Specialization for Energy-Efficient Computing},
  year={2012},
  volume={32},
  number={5},
  pages={38-51},
  doi={10.1109/MM.2012.51}}

@ARTICLE{amber,
  author={Feng, Kathleen and Kong, Taeyoung and Koul, Kalhan and Melchert, Jackson and Carsello, Alex and Liu, Qiaoyi and Nyengele, Gedeon and Strange, Maxwell and Zhang, Keyi and Nayak, Ankita and Setter, Jeff and Thomas, James and Sreedhar, Kavya and Chen, Po-Han and Bhagdikar, Nikhil and Myers, Zach A. and D’Agostino, Brandon and Joshi, Pranil and Richardson, Stephen and Torng, Christopher and Horowitz, Mark and Raina, Priyanka},
  journal={IEEE Journal of Solid-State Circuits}, 
  title={{Amber}: A 16-nm System-on-Chip With a Coarse-Grained Reconfigurable Array for Flexible Acceleration of Dense Linear Algebra},
  year={2024},
  volume={59},
  number={3},
  pages={947-959},
  doi={10.1109/JSSC.2023.3313116}}

@INPROCEEDINGS{freepdk45,
  author={Stine, James E. and Castellanos, Ivan and Wood, Michael and Henson, Jeff and Love, Fred and Davis, W. Rhett and Franzon, Paul D. and Bucher, Michael and Basavarajaiah, Sunil and Oh, Julie and Jenkal, Ravi},
  booktitle={2007 IEEE International Conference on Microelectronic Systems Education (MSE'07)}, 
  title={{FreePDK}: An Open-Source Variation-Aware Design Kit},
  year={2007},
  volume={},
  number={},
  pages={173-174},
  doi={10.1109/MSE.2007.44}}

@misc{joules, url={https://www.cadence.com/en_US/home/tools/digital-design-and-signoff/power-analysis/joules-rtl-power-solution.html}, author={{Cadence Design Systems, Inc.}}, abstractNote={Cadence Joules RTL Power Solution delivers time-based RTL power analysis while still providing high-quality estimates of gates and wires.}, language={en},year={2025}}

@inproceedings{accelwattch,
author = {Kandiah, Vijay and Peverelle, Scott and Khairy, Mahmoud and Pan, Junrui and Manjunath, Amogh and Rogers, Timothy G. and Aamodt, Tor M. and Hardavellas, Nikos},
title = {{AccelWattch}: A Power Modeling Framework for Modern {GPUs}},
year = {2021},
isbn = {9781450385572},
publisher = {Association for Computing Machinery},
address = {New York, NY, USA},
url = {https://doi.org/10.1145/3466752.3480063},
doi = {10.1145/3466752.3480063},
booktitle = {MICRO-54: 54th Annual IEEE/ACM International Symposium on Microarchitecture},
pages = {738–753},
numpages = {16},
location = {Virtual Event, Greece},
series = {MICRO '21}
}

@inproceedings{gpu-rodinia,
  author={Che, Shuai and Boyer, Michael and Meng, Jiayuan and Tarjan, David and Sheaffer, Jeremy W. and Lee, Sang-Ha and Skadron, Kevin},
  booktitle={2009 IEEE International Symposium on Workload Characterization (IISWC)}, 
  title={{Rodinia}: A benchmark suite for heterogeneous computing},
  year={2009},
  volume={},
  number={},
  pages={44-54},
  doi={10.1109/IISWC.2009.5306797}}

@article{cgra_survey_1,
author = {Liu, Leibo and Zhu, Jianfeng and Li, Zhaoshi and Lu, Yanan and Deng, Yangdong and Han, Jie and Yin, Shouyi and Wei, Shaojun},
title = {A Survey of Coarse-Grained Reconfigurable Architecture and Design: Taxonomy, Challenges, and Applications},
year = {2019},
issue_date = {November 2020},
publisher = {Association for Computing Machinery},
address = {New York, NY, USA},
volume = {52},
number = {6},
issn = {0360-0300},
url = {https://doi.org/10.1145/3357375},
doi = {10.1145/3357375},
journal = {ACM Comput. Surv.},
month = oct,
articleno = {118},
numpages = {39}
}

@article{cgra_survey_2,
  author={Podobas, Artur and Sano, Kentaro and Matsuoka, Satoshi},
  journal={IEEE Access}, 
  title={A Survey on Coarse-Grained Reconfigurable Architectures From a Performance Perspective}, 
  year={2020},
  volume={8},
  number={},
  pages={146719-146743},
  doi={10.1109/ACCESS.2020.3012084}}

@article{end_of_moore,
  author={Theis, Thomas N. and Wong, H.-S. Philip},
  journal={Computing in Science \& Engineering}, 
  title={The End of {Moore}'s Law: A New Beginning for Information Technology},
  year={2017},
  volume={19},
  number={2},
  pages={41-50},
  doi={10.1109/MCSE.2017.29}}

@article{inefficiency_in_gpc,
author = {Hameed, Rehan and Qadeer, Wajahat and Wachs, Megan and Azizi, Omid and Solomatnikov, Alex and Lee, Benjamin C. and Richardson, Stephen and Kozyrakis, Christos and Horowitz, Mark},
title = {Understanding sources of inefficiency in general-purpose chips},
year = {2010},
isbn = {9781450300537},
publisher = {Association for Computing Machinery},
address = {New York, NY, USA},
url = {https://doi.org/10.1145/1815961.1815968},
doi = {10.1145/1815961.1815968},
booktitle = {Proceedings of the 37th Annual International Symposium on Computer Architecture},
pages = {37–47},
numpages = {11},
location = {Saint-Malo, France},
series = {ISCA '10}
}

@inproceedings{nre_cost,
author = {Khazraee, Moein and Zhang, Lu and Vega, Luis and Taylor, Michael Bedford},
title = {{Moonwalk}: {NRE} Optimization in {ASIC} Clouds},
year = {2017},
isbn = {9781450344654},
publisher = {Association for Computing Machinery},
address = {New York, NY, USA},
url = {https://doi.org/10.1145/3037697.3037749},
doi = {10.1145/3037697.3037749},
booktitle = {Proceedings of the Twenty-Second International Conference on Architectural Support for Programming Languages and Operating Systems},
pages = {511–526},
numpages = {16},
location = {Xi'an, China},
series = {ASPLOS '17}
}

@book{reconfigurable_computing, address={San Francisco}, series={Systems on Silicon Ser}, title={Reconfigurable Computing: The Theory and Practice of {FPGA}-Based Computation}, ISBN={978-0-12-370522-8}, publisher={Morgan Kaufmann}, editor={Hauck, Scott and DeHon, Andr\'{e}}, year={2007}, collection={Systems on Silicon Ser}, language={en} }

@inproceedings{opencgra,
  author={Tan, Cheng and Xie, Chenhao and Li, Ang and Barker, Kevin J. and Tumeo, Antonino},
  booktitle={2020 IEEE 38th International Conference on Computer Design (ICCD)}, 
  title={{OpenCGRA}: An Open-Source Unified Framework for Modeling, Testing, and Evaluating {CGRAs}},
  year={2020},
  volume={},
  number={},
  pages={381-388},
  doi={10.1109/ICCD50377.2020.00070}}

@inproceedings{snafu,
  author={Gobieski, Graham and Atli, Ahmet Oguz and Mai, Kenneth and Lucia, Brandon and Beckmann, Nathan},
  booktitle={2021 ACM/IEEE 48th Annual International Symposium on Computer Architecture (ISCA)}, 
  title={{SNAFU}: An Ultra-Low-Power, Energy-Minimal {CGRA}-Generation Framework and Architecture},
  year={2021},
  volume={},
  number={},
  pages={1027-1040},
  doi={10.1109/ISCA52012.2021.00084}}

@inproceedings{hycube,
  author={Karunaratne, Manupa and Mohite, Aditi Kulkarni and Mitra, Tulika and Peh, Li-Shiuan},
  booktitle={2017 54th ACM/EDAC/IEEE Design Automation Conference (DAC)}, 
  title={{HyCUBE}: A {CGRA} with reconfigurable single-cycle multi-hop interconnect},
  year={2017},
  volume={},
  number={},
  pages={1-6},
  doi={10.1145/3061639.3062262}}

@inproceedings{riptide,
  author={Gobieski, Graham and Ghosh, Souradip and Heule, Marijn and Mowry, Todd and Nowatzki, Tony and Beckmann, Nathan and Lucia, Brandon},
  booktitle={2022 55th IEEE/ACM International Symposium on Microarchitecture (MICRO)}, 
  title={{RipTide}: A Programmable, Energy-Minimal Dataflow Compiler and Architecture},
  year={2022},
  volume={},
  number={},
  pages={546-564},
  doi={10.1109/MICRO56248.2022.00046}}

@inproceedings{plasticine,
  author={Prabhakar, Raghu and Zhang, Yaqi and Koeplinger, David and Feldman, Matt and Zhao, Tian and Hadjis, Stefan and Pedram, Ardavan and Kozyrakis, Christos and Olukotun, Kunle},
  booktitle={2017 ACM/IEEE 44th Annual International Symposium on Computer Architecture (ISCA)}, 
  title={{Plasticine}: A reconfigurable architecture for parallel patterns},
  year={2017},
  volume={},
  number={},
  pages={389-402},
  doi={10.1145/3079856.3080256}}

@article{sgmf,
author = {Voitsechov, Dani and Etsion, Yoav},
title = {Single-graph multiple flows: energy efficient design alternative for {GPGPUs}},
year = {2014},
isbn = {9781479943944},
publisher = {IEEE Press},
booktitle = {Proceeding of the 41st Annual International Symposium on Computer Architecuture},
pages = {205–216},
numpages = {12},
location = {Minneapolis, Minnesota, USA},
series = {ISCA '14}
}

@inproceedings{vgiw,
  author={Voitsechov, Dani and Etsion, Yoav},
  booktitle={2015 48th Annual IEEE/ACM International Symposium on Microarchitecture (MICRO)}, 
  title={Control flow coalescing on a hybrid dataflow/von {Neumann} {GPGPU}},
  year={2015},
  volume={},
  number={},
  pages={216-227},
  doi={10.1145/2830772.2830817}}

@article{cuda,
author = {Nickolls, John and Buck, Ian and Garland, Michael and Skadron, Kevin},
title = {Scalable Parallel Programming with {CUDA}: Is {CUDA} the Parallel Programming Model that Application Developers Have Been Waiting for?},
year = {2008},
issue_date = {March/April 2008},
publisher = {Association for Computing Machinery},
address = {New York, NY, USA},
volume = {6},
number = {2},
issn = {1542-7730},
url = {https://doi.org/10.1145/1365490.1365500},
doi = {10.1145/1365490.1365500},
journal = {Queue},
month = mar,
pages = {40-53},
numpages = {14}
}

@inproceedings{swizzle,
author = {Phothilimthana, Phitchaya Mangpo and Elliott, Archibald Samuel and Wang, An and Jangda, Abhinav and Hagedorn, Bastian and Barthels, Henrik and Kaufman, Samuel J. and Grover, Vinod and Torlak, Emina and Bodik, Rastislav},
title = {{Swizzle Inventor}: Data Movement Synthesis for {GPU} Kernels},
year = {2019},
isbn = {9781450362405},
publisher = {Association for Computing Machinery},
address = {New York, NY, USA},
url = {https://doi.org/10.1145/3297858.3304059},
doi = {10.1145/3297858.3304059},
booktitle = {Proceedings of the Twenty-Fourth International Conference on Architectural Support for Programming Languages and Operating Systems},
pages = {65–78},
numpages = {14},
location = {Providence, RI, USA},
series = {ASPLOS '19}
}

@techreport{turing,
  author       = {{NVIDIA Corporation}},
  title        = {{NVIDIA TURING GPU ARCHITECTURE}},
  howpublished = {Tech. Rep.},
  institution  = {NVIDIA},
  address      = {Santa Clara, CA},
  year         = {2018},
  note         = {\url{https://images.nvidia.com/aem-dam/en-zz/Solutions/design-visualization/technologies/turing-architecture/NVIDIA-Turing-Architecture-Whitepaper.pdf}},
}

@inproceedings{dmt-cgra,
  author={Voitsechov, Dani and Port, Oron and Etsion, Yoav},
  booktitle={2018 51st Annual IEEE/ACM International Symposium on Microarchitecture (MICRO)}, 
  title={{Inter-Thread Communication in Multithreaded, Reconfigurable Coarse-Grain Arrays}},
  year={2018},
  volume={},
  number={},
  pages={42-54},
  doi={10.1109/MICRO.2018.00013}}

@inproceedings{trips,
  author={Sankaralingam, K. and Nagarajan, R. and Haiming Liu and Changkyu Kim and Jaehyuk Huh and Burger, D. and Keckler, S.W. and Moore, C.R.},
  booktitle={30th Annual International Symposium on Computer Architecture, 2003. Proceedings.}, 
  title={Exploiting {ILP}, {TLP}, and {DLP} with the polymorphous {TRIPS} architecture},
  year={2003},
  volume={},
  number={},
  pages={422-433},
  doi={10.1109/ISCA.2003.1207019}}

@inproceedings{stream_dataflow,
  author={Nowatzki, Tony and Gangadhar, Vinay and Ardalani, Newsha and Sankaralingam, Karthikeyan},
  booktitle={2017 ACM/IEEE 44th Annual International Symposium on Computer Architecture (ISCA)}, 
  title={Stream-dataflow acceleration}, 
  year={2017},
  volume={},
  number={},
  pages={416-429},
  doi={10.1145/3079856.3080255}}

@inproceedings{wavescalar,
  author={Swanson, S. and Michelson, K. and Schwerin, A. and Oskin, M.},
  booktitle={Proceedings. 36th Annual IEEE/ACM International Symposium on Microarchitecture, 2003. MICRO-36.}, 
  title={{WaveScalar}},
  year={2003},
  volume={},
  number={},
  pages={291-302},
  doi={10.1109/MICRO.2003.1253203}}

@inproceedings{adres,
author="Mei, Bingfeng
and Vernalde, Serge
and Verkest, Diederik
and De Man, Hugo
and Lauwereins, Rudy",
editor="Y. K. Cheung, Peter
and Constantinides, George A.",
title="{ADRES}: An Architecture with Tightly Coupled {VLIW} Processor and Coarse-Grained Reconfigurable Matrix",
booktitle="Field Programmable Logic and Application",
year="2003",
publisher="Springer Berlin Heidelberg",
address="Berlin, Heidelberg",
pages="61--70",
isbn="978-3-540-45234-8"
}

@article{scale,
title = {Scaling equations for the accurate prediction of {CMOS} device performance from 180 nm to 7 nm},
journal = {Integration},
volume = {58},
pages = {74-81},
year = {2017},
issn = {0167-9260},
doi = {https://doi.org/10.1016/j.vlsi.2017.02.002},
url = {https://www.sciencedirect.com/science/article/pii/S0167926017300755},
author = {Aaron Stillmaker and Bevan Baas}
}

@inproceedings{cacti,
  author={Muralimanohar, Naveen and Balasubramonian, Rajeev and Jouppi, Norm},
  booktitle={40th Annual IEEE/ACM International Symposium on Microarchitecture (MICRO 2007)}, 
  title={Optimizing {NUCA} Organizations and Wiring Alternatives for Large Caches with {CACTI} 6.0},
  year={2007},
  volume={},
  number={},
  pages={3-14},
  doi={10.1109/MICRO.2007.33}}

@manual{ptxisa88,
  title        = {Parallel Thread Execution (PTX) ISA Version 8.8},
  author       = {{NVIDIA Corporation}},
  year         = {2024},
  note         = {\url{https://docs.nvidia.com/cuda/parallel-thread-execution/}},
}

@misc{volta,
  author       = {{NVIDIA Corporation}},
  title        = {{NVIDIA TESLA V100 GPU ARCHITECTURE}},
  year         = {2017},
  note         = {White paper},
  howpublished = {\url{https://images.nvidia.com/content/volta-architecture/pdf/volta-architecture-whitepaper.pdf}}
}

@misc{pascal, author       = {{NVIDIA Corporation}}, title = {{NVIDIA Tesla P100}}, howpublished ={\url{https://images.nvidia.com/content/pdf/tesla/whitepaper/pascal-architecture-whitepaper.pdf}}, note         = {White paper}, year         = {2016}, language={en-us} }

@misc{a100,
  author       = {{NVIDIA Corporation}},
  title        = {{NVIDIA A100 Tensor Core GPU Architecture}},
  year         = {2020},
  note         = {White paper},
  howpublished = {\url{https://images.nvidia.com/aem-dam/en-zz/Solutions/data-center/nvidia-ampere-architecture-whitepaper.pdf}}
}

@misc{3070,
  author       = {{NVIDIA Corporation}},
  title        = {{NVIDIA Ampere GA102 GPU Architecture}},
  year         = {2020},
  note         = {White paper},
  howpublished = {\url{https://www.nvidia.com/content/PDF/nvidia-ampere-ga-102-gpu-architecture-whitepaper-v2.pdf}}
}

@misc{h100,
  author       = {{NVIDIA Corporation}},
  title        = {{NVIDIA H100 Tensor Core GPU Architecture}},
  year         = {2022},
  note         = {White paper},
  howpublished = {\url{https://resources.nvidia.com/en-us-hopper-architecture/nvidia-h100-tensor-c}}
}

@misc{design_space,
      title={Automated Design Space Exploration of {CGRA} Processing Element Architectures using Frequent Subgraph Analysis},
      author={Jackson Melchert and Kathleen Feng and Caleb Donovick and Ross Daly and Clark Barrett and Mark Horowitz and Pat Hanrahan and Priyanka Raina},
      year={2021},
      eprint={2104.14155},
      archivePrefix={arXiv},
      primaryClass={cs.AR},
      url={https://arxiv.org/abs/2104.14155}, 
}

@inproceedings{cgra_me,
  author={Chin, S. Alexander and Sakamoto, Noriaki and Rui, Allan and Zhao, Jim and Kim, Jin Hee and Hara-Azumi, Yuko and Anderson, Jason},
  booktitle={2017 IEEE 28th International Conference on Application-specific Systems, Architectures and Processors (ASAP)}, 
  title={{CGRA-ME}: A unified framework for {CGRA} modelling and exploration},
  year={2017},
  volume={},
  number={},
  pages={184-189},
  doi={10.1109/ASAP.2017.7995277}}

@inproceedings{vecpac,
  author={Tan, Cheng and Patil, Deepak and Tumeo, Antonino and Weisz, Gabriel and Reinhardt, Steve and Zhang, Jeff},
  booktitle={2023 IEEE/ACM International Conference on Computer Aided Design (ICCAD)}, 
  title={{VecPAC}: A Vectorizable and Precision-Aware {CGRA}},
  year={2023},
  volume={},
  number={},
  pages={1-9},
  doi={10.1109/ICCAD57390.2023.10323910}}

@article{tag_token_1,
  author={Arvind and Nikhil, R.S.},
  journal={IEEE Transactions on Computers}, 
  title={Executing a program on the {MIT} tagged-token dataflow architecture},
  year={1990},
  volume={39},
  number={3},
  pages={300-318},
  doi={10.1109/12.48862}}

@inproceedings{tag_token_2,
  author={Watson, Ian and Gurd, John},
  booktitle={1979 International Workshop on Managing Requirements Knowledge (MARK)}, 
  title={A prototype data flow computer with token labelling}, 
  year={1979},
  volume={},
  number={},
  pages={623-628},
  doi={10.1109/MARK.1979.8817164}}

@misc{nvcc,
  author = {{NVIDIA Corporation}},
  title = {{CUDA Toolkit Documentation: NVCC, the CUDA Compiler Driver}},
  year = {2025},
  month = {June},
  howpublished = {\url{https://docs.nvidia.com/cuda/cuda-compiler-driver-nvcc/index.html}},
  note = {Accessed: 2025-07-30}
}

@inproceedings{mshr,
author = {Kroft, David},
title = {Lockup-free instruction fetch/prefetch cache organization},
year = {1981},
publisher = {IEEE Computer Society Press},
address = {Washington, DC, USA},
booktitle = {Proceedings of the 8th Annual Symposium on Computer Architecture},
pages = {81–87},
numpages = {7},
location = {Minneapolis, Minnesota, USA},
series = {ISCA '81}
}

@INPROCEEDINGS{scalesim-v3,
  author={Raj, Ritik and Banerjee, Sarbartha and Chandra, Nikhil and Wan, Zishen and Tong, Jianming and Samajdhar, Ananda and Krishna, Tushar},
  booktitle={2025 IEEE International Symposium on Performance Analysis of Systems and Software (ISPASS)}, 
  title={{SCALE-Sim} {V3}: a Modular Cycle-Accurate Systolic Accelerator Simulator for End-To-End System Analysis},
  year={2025},
  volume={},
  number={},
  pages={186-200},
  doi={10.1109/ISPASS64960.2025.00026}}

@inproceedings{gpu_power_model_2,
author = {Antepara, Oscar and Zhao, Zhengji and Austin, Brian and Ding, Nan and Oliker, Leonid and Wright, Nicholas J. and Williams, Samuel},
title = {Benchmark-driven Models for Energy Analysis and Attribution of {GPU}-Accelerated Supercomputing},
year = {2025},
isbn = {9798400714665},
publisher = {Association for Computing Machinery},
address = {New York, NY, USA},
url = {https://doi.org/10.1145/3712285.3759815},
doi = {10.1145/3712285.3759815},
booktitle = {Proceedings of the International Conference for High Performance Computing, Networking, Storage and Analysis},
pages = {888–904},
numpages = {17},
location = {
},
series = {SC '25}
}

@mastersthesis{fused_alu,
  author  = {Bruintjes, Tom M.},
  title   = {Design of a Fused Multiply-Add Floating-Point and Integer Datapath},
  school  = {University of Twente},
  year    = {2011},
  type    = {Master's thesis},
  address = {Enschede, The Netherlands},
  url     = {https://essay.utwente.nl/fileshare/file/61055/MSc_TM_Bruintjes_CAES_ASCI.pdf#page=121.00}
}

@inproceedings{pe_eval,
  author={Weng, Jian and Liu, Sihao and Wang, Zhengrong and Dadu, Vidushi and Nowatzki, Tony},
  booktitle={2020 IEEE International Symposium on High Performance Computer Architecture (HPCA)}, 
  title={A Hybrid Systolic-Dataflow Architecture for Inductive Matrix Algorithms}, 
  year={2020},
  volume={},
  number={},
  pages={703-716},
  doi={10.1109/HPCA47549.2020.00063}}

@misc{fritz_tu106_2019,
  author       = {Fritzchens Fritz},
  title        = {{Nvidia 12nm Turing TU106 GeForce RTX 2060 Die Shot}},
  year         = {2019},
  howpublished = {Flickr},
  url          = {https://www.flickr.com/photos/130561288@N04/46357922535},
  urldate      = {2026-03-28},
  note         = {CC0 1.0 / public domain}
}

@misc{fritz_tu116_2019,
  author       = {Fritzchens Fritz},
  title        = {{Nvidia 12nm Turing TU116 GeForce GTX 1660 Ti Die Shot}},
  year         = {2019},
  howpublished = {Flickr},
  url          = {https://www.flickr.com/photos/130561288@N04/47220396142},
  urldate      = {2026-03-28},
  note         = {CC0 1.0 / public domain}
}

@article{nvidia_gtx1660ti_tu116,
  author       = {Andrew Burnes},
  title        = {{GeForce GTX 1660 Ti's Advanced Shaders Accelerate Performance In The Latest Games}},
  journal      = {NVIDIA GeForce News},
  year         = {2019},
  month        = feb,
  day          = {25},
  url          = {https://www.nvidia.com/en-us/geforce/news/geforce-gtx-1660-ti-advanced-shaders-streaming-multiprocessor/},
  note         = {Accessed March 28, 2026}
}

\end{document}